\documentclass[journal=aesccq,manuscript=article]{achemso}

\usepackage{chemformula} 
\usepackage[T1]{fontenc} 
\usepackage{amsmath} 
\usepackage{amssymb}
\usepackage{placeins}
\usepackage{threeparttable}
\usepackage{titlesec}
\author{Beatrice M. Kulterer}
\affiliation{Center for Space and Habitability, University of Bern, Gesellschaftsstrasse 6, 3012 Bern, Switzerland}
\email{beatrice.kulterer@unibe.ch}
\author{Maria N. Drozdovskaya}
\affiliation{Center for Space and Habitability, University of Bern, Gesellschaftsstrasse 6, 3012 Bern, Switzerland}
\author{Stefano Antonellini}
\affiliation{Institut de Radioastronomie Millimétrique (IRAM), 300 rue de la piscine, F-38406
Saint-Martin d’Hères, France}
\author{Catherine Walsh}
\affiliation{School of Physics and Astronomy, University of Leeds, Leeds, LS2 9JT, UK}
\author{Tom J. Millar}
\affiliation{Astrophysics Research Centre, School of Mathematics and Physics, Queen’s University Belfast, University Road, Belfast
BT7 1NN, UK}

\title[]
  {Fevering Interstellar Ices Have More \ch{CH3OD}}

\abbreviations{IR,NMR,UV}

\begin{document}

\begin{tocentry}

\includegraphics[width=0.8\textwidth]{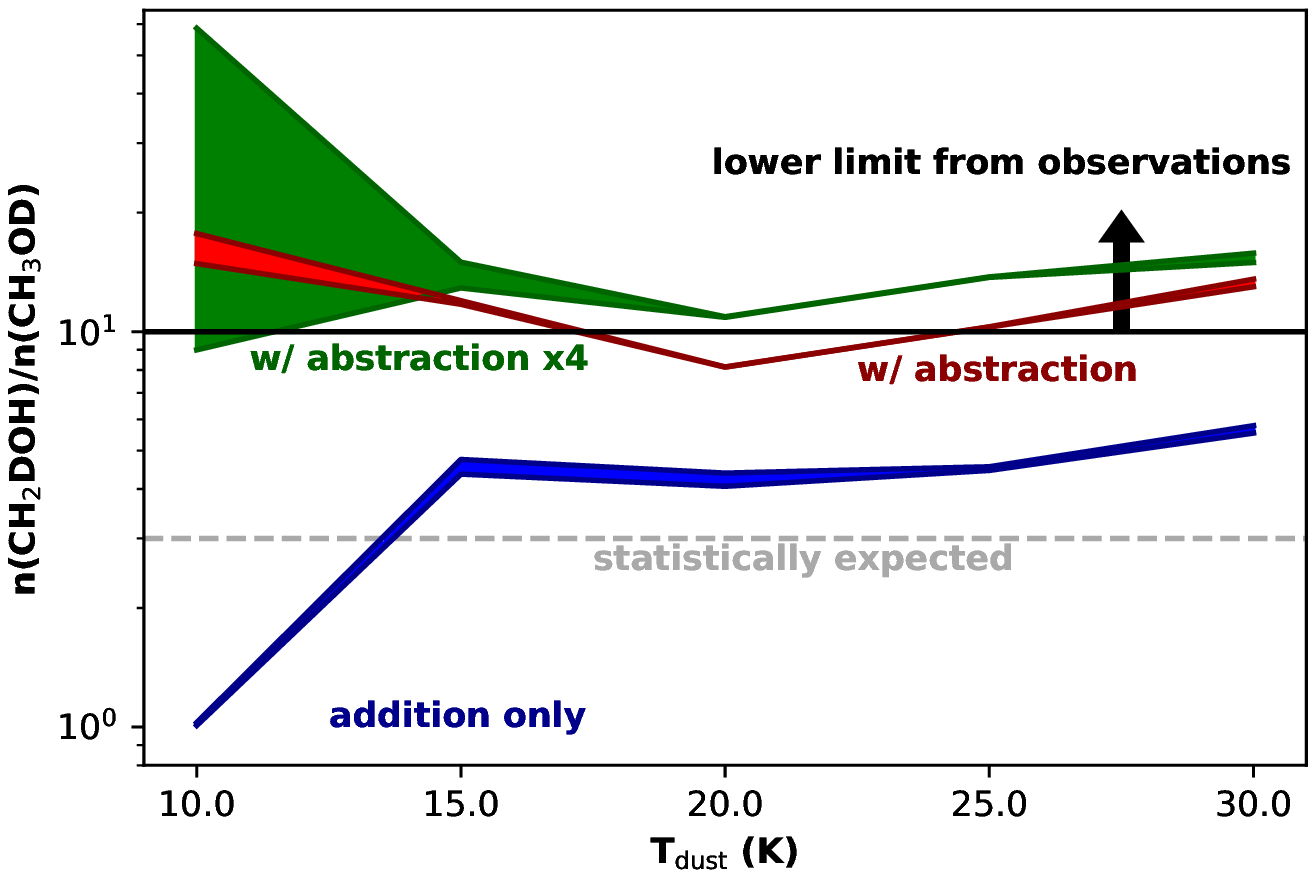}

\end{tocentry}

\begin{abstract}
  \noindent Mono-deuterated methanol is thought to form during the prestellar core stage of star formation. Observed variations in the \ch{CH2DOH}/\ch{CH3OD} ratio suggest that its formation is strongly dependent on the surrounding cloud conditions. Thus, it is a potential tracer of the physical conditions before the onset of star formation.  A single-point physical model representative of a typical prestellar core is coupled to chemical models to investigate potential formation pathways towards deuterated methanol at the prestellar stage. Simple addition reactions of H and D are not able to reproduce observed abundances. The implementation of an experimentally verified abstraction scheme leads to the efficient formation of methyl-deuterated methanol, but lacks sufficient formation of hydroxy-deuterated methanol. \ch{CH3OD} is most likely formed at a later evolutionary stage, potentially from H-D exchange reactions in warm ices between HDO (and \ch{D2O}) and \ch{CH3OH}. The  \ch{CH2DOH}/\ch{CH3OD} ratio is not an appropriate tracer of the physical conditions during the prestellar stage, but might be better suited as a tracer of ice heating.
\end{abstract}

\section{Introduction}
Cold dense cores are the birth places of stars like our Sun and already show contracting motions and signs of chemical evolution. They provide a unique opportunity to investigate the chemical composition before the onset of gravitational collapse and the subsequent formation of a protostellar
system \cite{Crapsi05,Caselli12}. Characterized by low temperatures ($\sim$~10~K) and gas densities of a few $\times$~10$^4$~cm$^{-3}$ \cite{Benson89}, their density increases to $\geq$~10$^5$~cm$^{-3}$ over their lifetime of $\sim$~10$^5$--10$^6$~yr \cite{Enoch08}, as they evolve from starless to prestellar and subsequently to protostellar cores. Prestellar cores can be distinguished from starless cores by identifying a combination of properties such as a higher central \ch{H2} column density and pronounced CO depletion \cite{Crapsi05}. Despite their cold nature, many molecules have been detected in prestellar cores, which can provide insight on their physical conditions.
\newline 
\noindent Since the first detection of an interstellar molecule \cite{Swings37}, the current number of molecular detections in the interstellar medium (ISM) has exceeded 250 \cite{McGuire18,McGuire21}. A notable number of them are so-called complex organic molecules (COMs), which have been defined to consist of at least six atoms, one of them being carbon \cite{Herbst09}. Besides simple chemical species, prestellar cores also harbor COMs \cite{Bacmann12,Vastel14,JimenezSerra16}, which supports chemical complexity commencing before the onset of star formation. This establishes the need to explore potential formation pathways that are feasible under these prestellar core conditions. The starting point of this molecular complexity is an on-going point of discussion, methanol (\ch{CH3OH}) is a strong candidate. For instance, radicals produced from UV-photolysis of methanol can produce species with a higher degree of complexity \cite{Oberg09}. Also, studies investigating electron-induced reactions on methanol ice detected the formation of organic species with a larger degree of complexity than methanol \cite{Boyer16,Sullivan16,Schmidt21}. Moreover, experiments have demonstrated that the hydrogenation of CO ice can produce species like glycerol, \ch{HOCH2CH(OH)CH2OH}, with methanol as an intermediate \cite{Fedoseev17}. Grain-surface formation of methanol has been studied in numerous experimental and theoretical modelling studies \cite{Fuchs09,Watanabe02,Cuppen09,Vasyunin13} and is needed to explain the methanol abundances in the gas and in the solid state in dense cores.
\newline 
\noindent An important chemical process that takes place in prestellar cores is deuterium fractionation. The reaction \ch{H3+} + HD $\longrightarrow$ \ch{H2D+} + \ch{H2} + 230K cannot proceed backwards if the kinetic temperature is below $\sim$~20~K \cite{Watson74}, which enhances the elemental abundance of deuterium relative to hydrogen (=~A$_{\rm D}$) \cite{Caselli12} in molecules with respect to its average value in the local ISM of 2.0~$\pm$~0.1~$\times$~10$^{-5}$ \cite{Prodanovic10}. Deuterium atoms and molecules with an increased deuterium over hydrogen (D/H) ratio can accrete onto icy grains and therefore also increase the deuterium fractionation in grain-surface species. Further enhancement of this ratio occurs due to the longer residence time of D on grains compared to the lighter H atom \cite{Tielens83}. As the deuteration is most efficient at the cold, prestellar phase, deuterated molecules should also predominately form at this stage. Therefore, the D/H content of deuterated molecules should be tied closely to the underlying physical properties of the prestellar environment \cite{Caselli12}. Under the assumption that such molecules do not get reprocessed during the subsequent stages of star formation, the deuterated molecules provide an opportunity to constrain the physical parameters of their birth cloud if their formation pathways are known. Quantifying the D/H ratio in molecules across the various stages of star formation and comparing them with D/H ratios obtained from cometary volatiles in the Solar System can shed light on the interstellar origin of cometary material \cite{Drozdovskaya21}. Comets may be the sole remaining link between the mature Solar System and its natal cloud, because they are thought to contain the most pristine remaining materials from the time of formation of the Solar System \cite{BockeleeMorvan15}. 
\newline 
\noindent Given the key role in starting chemical complexity in space of the main methanol isotopologue, and its abundance and occurrence in a wide range of environments, its deuterated isotopologues are promising candidates to investigate if COMs can be inherited from the earliest stages of star formation. This can be conducted via theoretical modelling work and complementary observations across the star formation sequence. The mono-deuterated isotopologues, \ch{CH2DOH} and \ch{CH3OD}, are routinely observed around low-mass and high-mass protostars \cite{Parise06survey,Jorgensen18,Taquet19,Peng12,Neill13,Bogelund18}, and di- and tri-deuterated methanol have also been detected around low-mass protostars \cite{Parise02double,Parise04trip,Agundez19,Drozdovskaya22}. D/H ratios in the range of a few \% establish the idea that deuterated methanol observed in the warm gas around protostars is not formed in-situ, but inherited from prestellar stages, because high fractionation combined with abundant methanol formation is only achieveable in cold phases \cite{Geppert06,Millar89}.   
\newline 
\noindent However, the formation of mono-deuterated methanol itself is ill-constrained. If D substitution is equally efficient in both of the functional groups of methanol, \ch{CH3} and OH, the \ch{CH2DOH}/\ch{CH3OD} ratio should always be equal to three, because three H atoms can be replaced by a D atom in the methyl group, but only one H atom in the hydroxy group. Observations of mono-deuterated methanol paint a different picture. The \ch{CH2DOH}/\ch{CH3OD} ratio can exceed 10 in the case of low-mass protostars, whereas it approaches values around unity for high-mass protostars \cite{Ratajczak11,Bogelund18,Taquet19}. Thus, methyl-deuterated methanol (where the functional group \ch{CH3} is deuterated) is overabundant compared to hydroxy-deuterated methanol (where the functional group OH is deuterated) in the case of low-mass protostars. This could be explained if \ch{CH2DOH} and \ch{CH3OD} formation occurs via different formation pathways, which are sensitive to the physical conditions of the cloud environment. Therefore, the ratio can momentarily only be investigated in terms of theoretical modelling work. Testing the sensitivity of the \ch{CH2DOH}/\ch{CH3OD} ratio during the prestellar core stage is the main goal of the work presented here.
\newline 
\newline 
\noindent The formation of deuterated methanol and the ratio of its mono-deuterated isotopologues, \ch{CH2DOH}/\ch{CH3OD}, is explored with different formation schemes that have been deduced from laboratory experiments for a parameter range characteristic of the prestellar core stage. Section 2 gives an overview of the physical and chemical model, and the different chemical networks tested. Results are presented in Section 3 and discussed in Section 4. The conclusions are stated in Section 5.  
\section{Models}
\subsection{Physical model}
The physical model is representative of a typical prestellar core \cite{Drozdovskaya16,Kulterer20}. The core age of the fiducial model, t$_{\rm core}$, is assumed to be 3~$\times$~10$^5$~yr \cite{Kulterer20}. The dust and gas temperatures (T$_{\rm dust}$) are assumed to be the same, and values of 10, 15, 20, 25, and 30~K are considered. The values for the gas density n$_{\rm H}$~=~4~$\times$~10$^4$~cm$^{-3}$ and the cosmic ray ionization rate $\zeta _{\rm CR}$~=~5~$\times$~10$^{-17}$~s$^{-1}$ remain the same for all discussed modelling runs with the fiducial physical model. However, an extended parameter space study at 10~K is presented in Section 3.2.1, where gas densities of 10$^5$, 5~$\times$~10$^5$, 10$^6$, and 5~$\times$~10$^6$~cm$^{-3}$ are investigated. 
\newline
\noindent Molecular abundances are calculated with a single point model that ages under constant prestellar cloud conditions. The chemical calculations begin with the evolution of initial atomic abundances (Table \ref{tbl:iniatomabun}), so that the discussed molecular abundances can be used as the molecular budget of a system at the beginning of the collapse phase and to compare to observations of starless and prestellar cores. 
\noindent While the main focus of this work is the fiducial model (in boldface, Table \ref{tbl:physgrid}), core ages ranging from 10$^4$--10$^7$~yr for the mentioned temperature grid are run (Table \ref{tbl:physgrid}).
\begin{table}
  \caption{Initial atomic abundances at the start of the calculations relative to n$_{\rm H}$.}
  \label{tbl:iniatomabun}
  \begin{tabular}{ll}
    \hline
    Species  & n(X)  \\
    \hline
    H & 5.00~$\times$~10$^{-5}$ \\
    \ch{H2} & 5.00~$\times$~10$^{-1}$ \\
    He & 9.75~$\times$~10$^{-2}$ \\
    C & 1.40~$\times$~10$^{-4}$ \\
    O & 3.20~$\times$~10$^{-4}$ \\
    S & 8.00~$\times$~10$^{-8}$ \\
    Si & 8.00~$\times$~10$^{-9}$ \\
    P & 3.00~$\times$~10$^{-9}$ \\
    Cl & 4.00~$\times$~10$^{-9}$ \\
    F & 2.00~$\times$~10$^{-8}$ \\
    Fe & 3.00~$\times$~10$^{-9}$ \\
    Na & 2.00~$\times$~10$^{-9}$ \\
    Mg & 7.00~$\times$~10$^{-9}$ \\
    HD & 1.60~$\times$~10$^{-5}$ \\
    \hline
  \end{tabular}
\end{table}

\begin{table}[h]
    \caption{Investigated parameter space with the three chemical models.$^{\rm a}$}
    \label{tbl:physgrid}
    \begin{tabular}{ll}
    \hline 
    Parameter     &  Value \\
    \hline 
     t$_{\rm core}$ (yr)    & 10$^4$, (1, \textbf{3}, 5, 7.5) \textbf{$\times$ 10$^5$}, 10$^6$, 10$^7$ \\
     T$_{\rm dust,gas}$ (K) & \textbf{10}, 15, 20, 25, 30 \\
     n$_{\rm H}$ (cm$^{-3}$) & \textbf{4~$\times$~10$^4$}, 10$^5$, 5~$\times$~10$^5$, 10$^6$, 5~$\times$~10$^6$ \\
     $\zeta _{\rm CR}$ (s$^{-1}$) & 5~$\times$~10$^{-17}$ \\
     E$_{\rm A(CO+H)}$ (K) & 400, 900, 1~400, \textbf{2~500} \\
     \hline 
    \end{tabular}
    \begin{tablenotes}
    \item[*] $^{\rm a}$ If multiple values are given per parameter, the default value used in the fiducial model is indicated in boldface.
    \end{tablenotes}
\end{table}

\subsection{Chemical model}
Here, a short description of the chemical model is given. Details can be found in Walsh et al. (2014), Drozdovskaya et al. (2014, 2016), and Kulterer et al. (2020) \cite{Walsh14,Drozdovskaya14,Drozdovskaya16,Kulterer20} and the references therein. The model considers gaseous and solid phases, the latter refers to icy mantles covering the dust grains. The bulk of the icy mantle and the surface layers are treated as one phase. However, only the upper two monolayers of the surface are chemically active \cite{Drozdovskaya16}. The chemical network is based on the deuterated version of the \textsc{Rate12} release of the \textsc{umist} Database for Astrochemistry (UDfA; \cite{McElroy13} Antonellini et al. in prep.). This base network mainly considers mono-deuterated species. COMs with a degree of complexity up to acetaldehyde (\ch{CH3CHO}), acetone (\ch{CH3COCH3}), dimethyl ether (\ch{CH3OCH3}), ethanol (\ch{C2H5OH)}, methyl formate (\ch{HCOOCH3}), and propene (\ch{CH3CHCH2}) are included. 
\noindent Gas-phase and grain-surface reaction rates are calculated via the rate equation approach \cite{Hasegawa92,Hasegawa93}. The parametrization for quantum tunneling is taken from the same authors \cite{Hasegawa92,Hasegawa93}. Gas-phase reaction types include typical two-body processes such as neutral-neutral or ion-molecule reactions, direct ionization by cosmic rays (CRs), and photodissociation and -ionization via internal and external UV photons. Interactions between the gas and the ice occur via adsorption, thermal desorption, photodesorption, and CR-induced spot heating. In contrast to the previous work carried out with this model, \cite{Drozdovskaya16,Kulterer20,Walsh14,Drozdovskaya14} reactive desorption is not included here. The Langmuir-Hinshelwood mechanism is implemented for the grain-surface chemical reactions \cite{Walsh14}. On the grains, H, D, \ch{H2}, and \ch{D2} are allowed to react via quantum tunneling with an assumed barrier thickness of 1~\r{A} \cite{Hasegawa92}, unless thermal hopping is faster. For all other species, only thermal hopping is considered. Photoreactions are also considered to occur on the grain surfaces. Experimental and theoretical work shows \cite{Bertin12} that photodesorption only occurs in the two uppermost monolayers of the ice. The inclusion of a coverage factor ensures that this criteria is fulfilled. The photodesorption yield of methanol is set to the value for prestellar cores of 1.5~$\times$~10$^{-5}$~molecule/photon presented in Bertin et al. (2016) \cite{Bertin16}. The grain radius is assumed to be uniform and set to 0.1~$\mu$m. Moreover, all grains are negatively charged and have a constant abundance of 2.2~$\times$~10$^{-12}$ relative to the H nuclei number density (n$_{\rm H}$~=~n(H)+2~$\times$~n(\ch{H2})). This is equivalent to a gas-to-dust mass ratio of $\sim$~100 \cite{Walsh14}. This approach is justified, as neutral grains will only dominate in the densest regions of protoplanetary disks, if the electron fraction is less than the grain fractional abundance \cite{Umebayashi09}. The diffusion energy, E$_{\rm diff}$, is set to 0.3~$\times$~E$_{\rm des}$ for all reactions. The desorption energy, E$_{\rm des}$, accounts for the unique binding energy to the grain surface of a species \cite{Walsh14}. The binding energies of the main and deuterated isotopologues are assumed to be equal \cite{Taquet13}.   
\subsubsection{Model 1}
The chemical network, denoted as Model 1, includes $\sim$ 29~200 reactions and 1~170 species. With the exception of \ch{HD2+}, \ch{D3+}, \ch{D2}, and \ch{D2O}, only mono-deuterated species are included, even if multiply deuterated isotopologues could theoretically be formed.
The main methanol isotopologue (\ch{CH3OH}) in the solid phase predominately forms via subsequent addition of hydrogen atoms to CO in this network. Laboratory experiments \cite{Fuchs09,Watanabe02} have shown that methanol formation occurs efficiently via the hydrogenation of CO ices at temperatures in the range of $\sim$~10--12~K  with \ch{H2CO} as an intermediate:
\begin{equation}
    \rm CO \longrightarrow HCO \longrightarrow \ch{H2CO} \longrightarrow \ch{CH3O},\ch{CH2OH} \longrightarrow \ch{CH3OH}.  
    \label{eqn:S1}
\end{equation}
These experiments have confirmed that the majority of methanol that is observed in the ISM likely stems from this reaction pathway. Gas-phase formation of methanol occurs, but not efficiently enough to explain the observed abundances \cite{Geppert06}. Calculated activation barriers for the first and third hydrogenation steps of the scheme in Eq. \ref{eqn:S1} are large ($>$~2 000~K) \cite{Woon02}, therefore they occur via quantum tunneling. The activation barrier for the formation of \ch{CH2OH} has been shown to be more than a factor of two higher \cite{Woon02,Osamura05} and therefore the formation of \ch{CH3O} is more likely. The activation energies are adopted from the \textsc{kida} database \cite{Wakelam12} and are set to 2~500~K for HCO, 2~200~K for \ch{CH3O}, and 5~400~K for \ch{CH2OH} formation. 
\newline 
\noindent In Model 1, mono-deuterated methanol forms via one D addition that replaces one H addition at any of the four reaction steps. \ch{DCO} forms via CO + D and can then be hydrogenated to form singly deuterated formaldehyde, \ch{HDCO}. Additionally, \ch{HDCO} can also be produced via HCO + D. \ch{H2CO} can either be hydrogenated to \ch{CH3O} or \ch{CH2OH}, or deuterated to \ch{CH2OD} or \ch{CH2DO}. The latter is also formed via H addition to HDCO. Alternatively, HDCO can produce CHDOH upon H addition. The activation energies of the deuterated isotopologues of \ch{CH2OH} and \ch{CH3O} (\ch{CH2DO}, \ch{CH2OD}, and CHDOH) are assumed to be the same as their non-deuterated counterparts. The final step in the formation of mono-deuterated methanol now provides three possible channels for \ch{CH2DOH}, and two for \ch{CH3OD}. \ch{CH3OD} can form from either \ch{CH2OD} or \ch{CH3O}. For the last step of \ch{CH2DOH} formation, addition reactions to \ch{CH2OH}, \ch{CH2DO}, and CHDOH are possible. All of the five intermediates are solely considered as grain-surface species, therefore they cannot be lost to the gas phase. This is based on findings that the time scales, under which hydrogenation and grain-surface accretion compared to those for desorption processes occur, do not favor a surface chemistry origin of gaseous HCO and \ch{CH3O} \cite{Bacmann16}.

\subsubsection{Model 2A}

Model 2A is an extension of Model 1 to investigate abstraction and substitution reactions via H and D along the pathway to form (multiply deuterated) methanol based on experimental findings \cite{Nagaoka05,Watanabe06,Nagaoka07,Hidaka09}. Note that only laboratory-verified abstraction routes by Hidaka and collaborators \cite{Hidaka09} and reactions with deuterated analogues of those specific abstraction routes have been included in the model rather than all possible abstraction routes. Therefore, the chemical network underlying Model 1 has been extended to include multiply deuterated species following practices common for the deuteration of chemical networks \cite{Millar89,Roberts03}. Multiply deuterated species have only been added, if their non- or mono-deuterated counterparts are directly involved in reactions with methanol in either Model 1; or if it is a species with deuteration already accounted for explicitly in the network presented in Antonellini et al. in prep. Hence, methanol is the only COM in this network that is considered to have multiply deuterated isotopologues, but multiply deuterated isotopologues of, e.g., \ch{NH3} and \ch{CH4} are still included. The final network consists of $\sim$~36~300 reactions and $\sim$~1~258 species. 
\newline
\noindent Five reaction schemes are included that lead to the formation of multiply deuterated methanol. Besides H and D addition (Section 2.2.1, Eq. 1), four substitution schemes are considered. They have been derived from experiments on amorphous solid water (ASW) in the temperature range of 10--30~K under simultaneous exposure to H and D atoms with a D/H ratio of 0.1 \cite{Nagaoka05,Nagaoka07,Hidaka09}. Therefore, the kinetic isotope effect (KIE) is set to 0.1 for all reaction rates in the scheme unless a relative reaction has been derived from the experiments. The KIE defines the change that occurs in a reaction rate if an atom is replaced by an isotope. The key reactions of all schemes are summarized in Table \ref{tbl:Hidakascheme}.  Triply deuterated methanol forms via the repeated formation of a hydroxymethyl radical as a result of H abstraction from methanol followed by D addition to the newly formed radical, i.e.:
\begin{equation}
    \rm \ch{CH3OH} + D \longrightarrow \ch{CH2OH} + HD
\end{equation}
\begin{equation}
    \rm \ch{CH2OH} + D \longrightarrow \ch{CH2DOH}.
\end{equation}
This process has been quantified in the laboratory \cite{Nagaoka05} and leads to the following sequence, S2:
\begin{equation}
    \rm \ch{CH3OH} \longrightarrow \ch{CH2DOH} \longrightarrow \ch{CHD2OH} \longrightarrow \ch{CD3OH}.
    \label{eqn:S2}
\end{equation}
Note that Nagaoka et al. \cite{Nagaoka05} do not observe deuteration of the hydroxy group in their experiments. The production of the -OD functional group is excluded up to the sensitivity of the instruments.
\newline 
\noindent The remaining three substitution schemes in the network are derived from experiments of the exposure of formaldehyde to H and D atoms \cite{Hidaka09}. The third scheme, S3, leads to the formation of di-deuterated formaldehyde and \ch{CD3OD} upon D exposure of \ch{H2CO}, and additionally forms HCO and DCO as intermediates:
\begin{equation}
    \rm \ch{H2CO} \longrightarrow \ch{HDCO} \longrightarrow \ch{D2CO} \longrightarrow \ch{CD3OD}.
    \label{eqn:S3}
\end{equation}
Furthermore, H exposure of \ch{D2CO} proceeds to form \ch{H2CO}, which then prompts \ch{CH3OH} production in a fourth scheme considered in this model (S4):
\begin{equation}
    \rm \ch{D2CO} \longrightarrow \ch{HDCO} \longrightarrow \ch{H2CO} \longrightarrow \ch{CH3OH}.
    \label{eqn:S4}
\end{equation}
The fifth and final reaction scheme, S5, generates \ch{CHD2OH}, again starting from \ch{D2CO} + H:
\begin{equation}
    \rm \ch{D2CO} \longrightarrow \ch{CHD2O} \longrightarrow \ch{CHD2OH}.
    \label{eqn:S5}
\end{equation}
The reaction rates for the abstraction scheme in Model 2A as detailed in Table \ref{tbl:Hidakascheme} are taken from Hidaka et al. 2009 \cite{Hidaka09}, which are given relative to the experimentally obtained reaction rate for CO + H on ASW at 15~K \cite{Hidaka04}. 
Besides the formation of \ch{CD3OD} \cite{Hidaka09} in the reaction sequence S3, deuteration of the hydroxy group was not seen in the experiments \cite{Nagaoka05,Nagaoka07,Hidaka09}. However, deuteration in the hydroxy group still proceeds via subsequent H and D addition analogous to S1 in this network. Furthermore, deuteration of the methyl group occurs via addition reactions, as in Model 1. Therefore, \ch{CH3OH} and all its deuterated isotopologues are included in the network. Note that no abstraction reactions from \ch{CD3OH} and \ch{CD3OD} are included in the abstraction scheme by Hidaka et al. \cite{Hidaka09}. Moreover, only reactions described in S1-5 are considered for the remaining deuterated isotopologues of methanol. Hence, no additional hydrogenation and deuteration steps to form more complex species directly from methanol are included in this network. Again, all of the intermediate radicals only exist as grain-surface species in this model. Following the surface reaction schemes presented by Nagaoka et al. \cite{Nagaoka07} and Hidaka et al. \cite{Hidaka09} the experiments solely find \ch{CX3O}, where X is either H or D, upon H and D addition to (deuterated) formaldehyde. Therefore, the relative reactions rates by the experiments are only used for reactions that involve intermediates of the type \ch{CX3O}. The barrier for \ch{CX2OX} formation from H and D addition to (deuterated) formaldehyde is set to 2~500~K. 
\begin{table}[h]
    \caption{Abstraction scheme towards the formation of deuterated methanol on grain surfaces.$^{\rm a}$}
    \label{tbl:Hidakascheme}
    \begin{tabular}{lllll}
    \hline 
     Reacting species    & & Produced species & Rate$^{\rm b}$ & Sequence$^{\rm c}$ \\
     \hline 
     CO + H & $\longrightarrow$ & HCO & 1.000 & S1 \\
     CO + D & $\longrightarrow$ & DCO & 0.100  & S1 \\
     \ch{H2CO} + H & $\longrightarrow$ & \ch{CH3O} & 0.500 & S1 \\
    \ch{H2CO} + H & $\longrightarrow$ & HCO + \ch{H2} & 0.000 & \textit{S2}$^{\rm d}$ \\
     \ch{H2CO} + D & $\longrightarrow$ & \ch{CH2DO} & 0.050 & S3 \\
     \ch{H2CO} + D & $\longrightarrow$ & HCO + HD & 0.485 & S3 \\
     \ch{H2CO} + D & $\longrightarrow$ & HDCO + H & 0.485 & S3 \\
     HDCO + H & $\longrightarrow$ & \ch{CH2DO} & 0.580 & S4 \\
     HDCO + H & $\longrightarrow$ & HCO + HD & 0.380 & S4 \\
     HDCO + H & $\longrightarrow$ & DCO + \ch{H2} & 0.000 & \textit{S4} \\
     HDCO + D & $\longrightarrow$ & \ch{CHD2O} & 0.058 & \textit{S1} \\
     HDCO + D & $\longrightarrow$ & HCO + \ch{D2} & 0.000 & \textit{S3} \\
     HDCO + D & $\longrightarrow$ & DCO + HD & 0.485 & S3 \\
     HDCO + D & $\longrightarrow$ & \ch{D2CO} + H & 0.485 & S3 \\
     \ch{D2CO} + H & $\longrightarrow$ &  \ch{CHD2O} & 0.660 & S5 \\
     \ch{D2CO} + H & $\longrightarrow$ & DCO + HD & 0.380 & S4 \\
     \ch{D2CO} + D & $\longrightarrow$ & \ch{CD3O} & 0.066 & \textit{S1} \\
     \ch{D2CO} + D & $\longrightarrow$ & DCO + \ch{D2} & 0.000 & \textit{S3} \\
     \ch{CH3OH} + D & $\longrightarrow$ & \ch{CH2OH} + HD & 1.500 & S2 \\
     \ch{CH2DOH} + D & $\longrightarrow$ & CHDOH + HD & 1.000 & S2 \\
     \ch{CHD2OH} + D & $\longrightarrow$ & \ch{CD2OH} + HD & 0.780 & S2 \\
     \ch{CH3OD} + D & $\longrightarrow$ & \ch{CH2OD} + HD & 1.500 & \textit{S2} \\
     \ch{CH2DOD} + D & $\longrightarrow$ & \ch{CHDOD} + HD & 1.000 & \textit{S2} \\
     \ch{CHD2OD} + D & $\longrightarrow$ & \ch{CD2OD} + HD & 0.780 & \textit{S2} \\
     \hline 
    \end{tabular}
    \begin{tablenotes}
    \item[*] $^{\rm a}$ as proposed in Hidaka et al. 2009 \cite{Hidaka09}; $^{\rm b}$ ``Rate'' in column 3 refers to the relative reaction rate that was applied to the reaction rate of CO + H $\longrightarrow$ HCO to obtain the reaction rates of the abstraction scheme; $^{\rm c}$ ``Sequence'' indicates from which of the proposed formation pathways the reaction was taken. The formation schemes can be found in the text as Eq. \ref{eqn:S1} for S1, Eq. \ref{eqn:S2} for S2, Eq. \ref{eqn:S3} for S3, and Eq. \ref{eqn:S4} for S4; $^{\rm d}$ Italic letters indicate, that the reaction is not explicitly part of the surface reaction network in Hidaka et al. 2009, but created analogously to a reaction in the denoted sequence.
    \end{tablenotes}
\end{table}

\FloatBarrier

\subsubsection{Model 2B}

Model 2B and Model 2A share the same underlying chemical network, but the relative rates for the abstraction scheme (Table \ref{tbl:Hidakascheme}) are changed: with the exception of CO + D $\longrightarrow$ DCO, the relative rates, which are scaled to the experimentally derived reaction rate of CO + H $\longrightarrow$ HCO at 15~K on ASW \cite{Hidaka04}, have been multiplied by an arbitrary factor of 4. This allows to check if the abundance of a molecule considered in the abstraction scheme is very sensitive to the chosen scaling factor or if the ratio of two molecules of interest is more dependent on the cloud age and dust temperature.  

\subsubsection{Investigation of the Reaction \ch{H2CO} + H with Model 2}

The reaction \ch{H2CO} + H has three product channels. Hydrogenation of \ch{H2CO} yields either \ch{CH3O} or \ch{CH2OH}, H abstraction produces HCO + \ch{H2}. So far, the first and third channel have been explored with relative rate constants scaled to the reaction rate of CO + H $\longrightarrow$ HCO from experiments \cite{Nagaoka07,Hidaka09}. The activation barrier for the second channel was set to 2~500~K \cite{Ruffle01,Garrod06,Drozdovskaya14}. However, as mentioned in Section 2.1, according to the \textsc{kida} database, the activation barrier for \ch{CH2OH} is 5~400~K, which was not included in Models 2A and 2B. Moreover, since the publications of the experimental findings, \cite{Nagaoka05,Watanabe06,Nagaoka07,Hidaka09} quantum chemical investigations of the three channels have been carried out \cite{Song17}. Therefore, two additional chemical models based on Model 2A that investigate the change in the \ch{CH2DOH}/\ch{CH3OD} ratio by varying the activation barriers of the first and third reactions with H in Eq. \ref{eqn:S1} are tested. The new implementations stem from quantum chemical calculations \cite{Song17}, and are taken from theoretical modelling work \cite{Taquet12} (T12 hereafter). 

\paragraph{Model 2A$\_$\ch{CX2OX}}

In contrast to Model 2A, in Model 2A$\_$\ch{CX2OX} the activation barriers for the reaction \ch{X2CO} + X $\longrightarrow$ \ch{CX2OX}, were X can be H or D, are changed to the value tabulated in the \textsc{kida} database \cite{Wakelam12}, 5~400~K. Moreover, five new addition reactions for \ch{CX2OX} formation are included in this version of the network that were not explicitly mentioned in the network derived from experimental work and; thus, not included in Model 2A \cite{Nagaoka07,Hidaka09} (see Supporting Information). Previous modelling work used activation barriers of 400, 900, and 1~400~K for the CO + H $\longrightarrow$ HCO reaction \cite{Taquet12}. Therefore, Model 2A$\_$\ch{CX2OX} is also used to investigate the impact of the variation of this activation barrier between 400 and 2~500~K on the \ch{CH2DOH}/\ch{CH3OD} ratio.

\paragraph{Model 2A$\_$Song}

The three product channels of the reaction \ch{H2CO} + H have recently been investigated on ASW with a quantum mechanics-molecular mechanics model by Song et al. (2017) \cite{Song17}. The tunneling rate constants given by Song et al. (2017) \cite{Song17} are not reliable below $\sim$~40~K, and therefore do not fit for the investigation carried out in this work, but their calculated activation barriers are implemented in Model 2A$\_$Song. For the main isotopologue reactions, the average of the values shown in Table 1 of Song et al. \cite{Song17} are adopted. Activation barriers for reactions involving D are taken from Table 2 from the same paper. Thus, relative reaction rates from the laboratory experiments by Hidaka and colleagues \cite{Hidaka09} are neglected for all three channels of \ch{X2CO} + X in this iteration of Model 2. As for Model 2A$\_$\ch{CX2OX}, the activation barrier of the CO + H $\longrightarrow$ HCO reaction varies from 400 to 2~500~K. The most distinct change in this model is the increase of the activation barrier of the abstraction reaction \ch{X2CO} + X $\longrightarrow$ XCO + \ch{X2}. The calculations of Song and peers \cite{Song17} show that the dissociation of \ch{H2CO} into HCO + \ch{H2} is efficiently suppressed by the ASW surface, a reaction channel that was deemed to contribute considerably to the formation of HCO and DCO in Models 2A and 2B (Sections 2.2.2 and 2.2.3). Depending on the number and position of D atoms in the reactions, E$_{\rm A}$ is in the 2~700--3~540~K range for the three product channels.

\section{Results}
First, results obtained with the fiducial model (Table \ref{tbl:physgrid}) are discussed here. Section 3.1 describes the abundances of (deuterated) methanol and its precursors for Models 1, 2A, and 2B. In Section 3.2, computations of the grain-surface \ch{CH2DOH}/\ch{CH3OD} ratio over the full t$_{\rm core}$-T$_{\rm dust}$ grid are detailed. Moreover, this ratio is explored for an extended parameter space at 10~K in Section 3.2.1. Caveats of the network are addressed in Section 3.3.

\subsection{Grain-surface abundances}

\begin{figure}
  \includegraphics[width=0.8\textwidth]{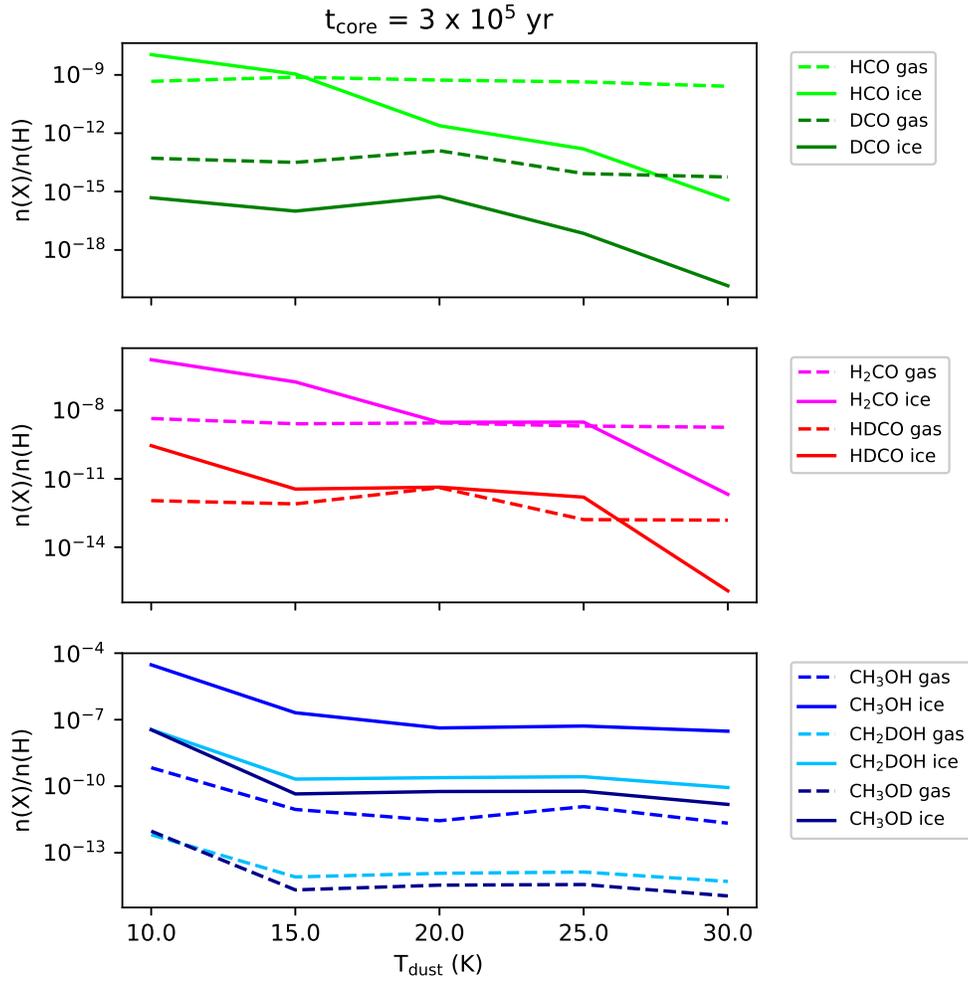}
  \caption{Abundances of key species towards the formation of mono-deuterated methanol as obtained with Model 1 for a core age of 3~$\times$~10$^5$~yr and dust temperatures in the range of 10--30~K. Gaseous species are depicted as dashed lines, solid lines refer to grain-surface species. \ch{HCO} (light green) and \ch{DCO} (green) are shown in the upper panel. The middle panel plots \ch{H2CO} (pink) and \ch{HDCO} (red). \ch{CH3OH} (blue) and its mono-deuterated isotopologues \ch{CH2DOH} (light blue) and \ch{CH3OD} (dark blue) are shown in the lower panel.}
  \label{fgr:3e5_Model1}
\end{figure}

\begin{figure}
  \includegraphics[width=0.8\textwidth]{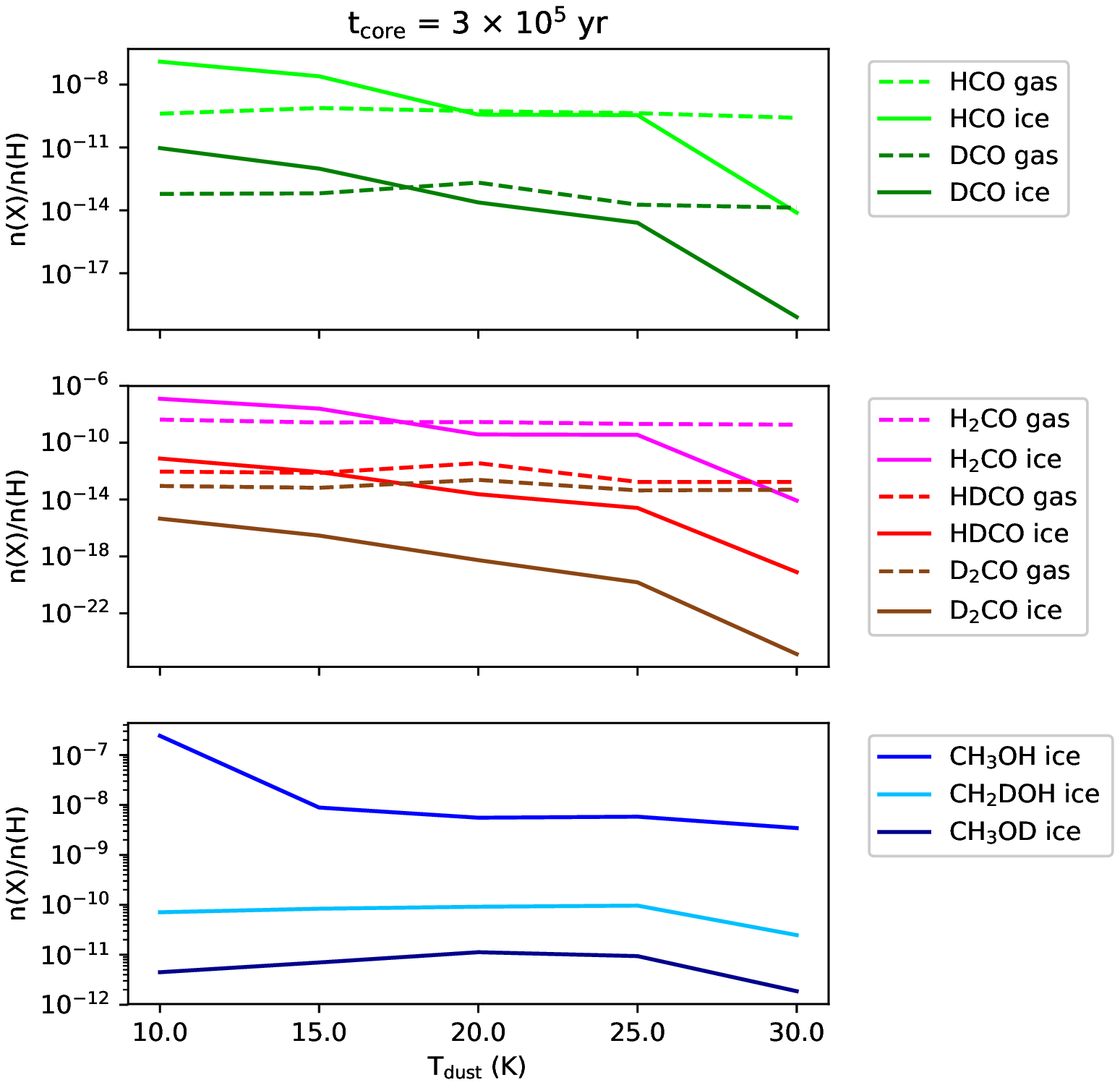}
  \caption{Abundances of key species towards the formation of deuterated methanol as obtained with Model 2A for a core age of 3~$\times$~10$^5$~yr and dust temperatures in the range of 10--30~K. Gaseous species are depicted as dashed lines, solid lines refer to grain-surface species. \ch{HCO} (light green) and \ch{DCO} (green) are shown in the upper panel. The middle panel plots \ch{H2CO} (pink), as well as \ch{HDCO} (red), and \ch{D2CO} (brown). \ch{CH3OH} (blue) and its mono-deuterated isotopologues, \ch{CH2DOH} (light blue) and \ch{CH3OD} (dark blue), are shown in the lower panel. Gas-phase abundances of (deuterated) methanol are $\leq$~10$^{-11}$ and therefore not included in the plot.}
  \label{fgr:3e5_Model2A}
\end{figure}

\begin{figure}
  \includegraphics[width=0.8\textwidth]{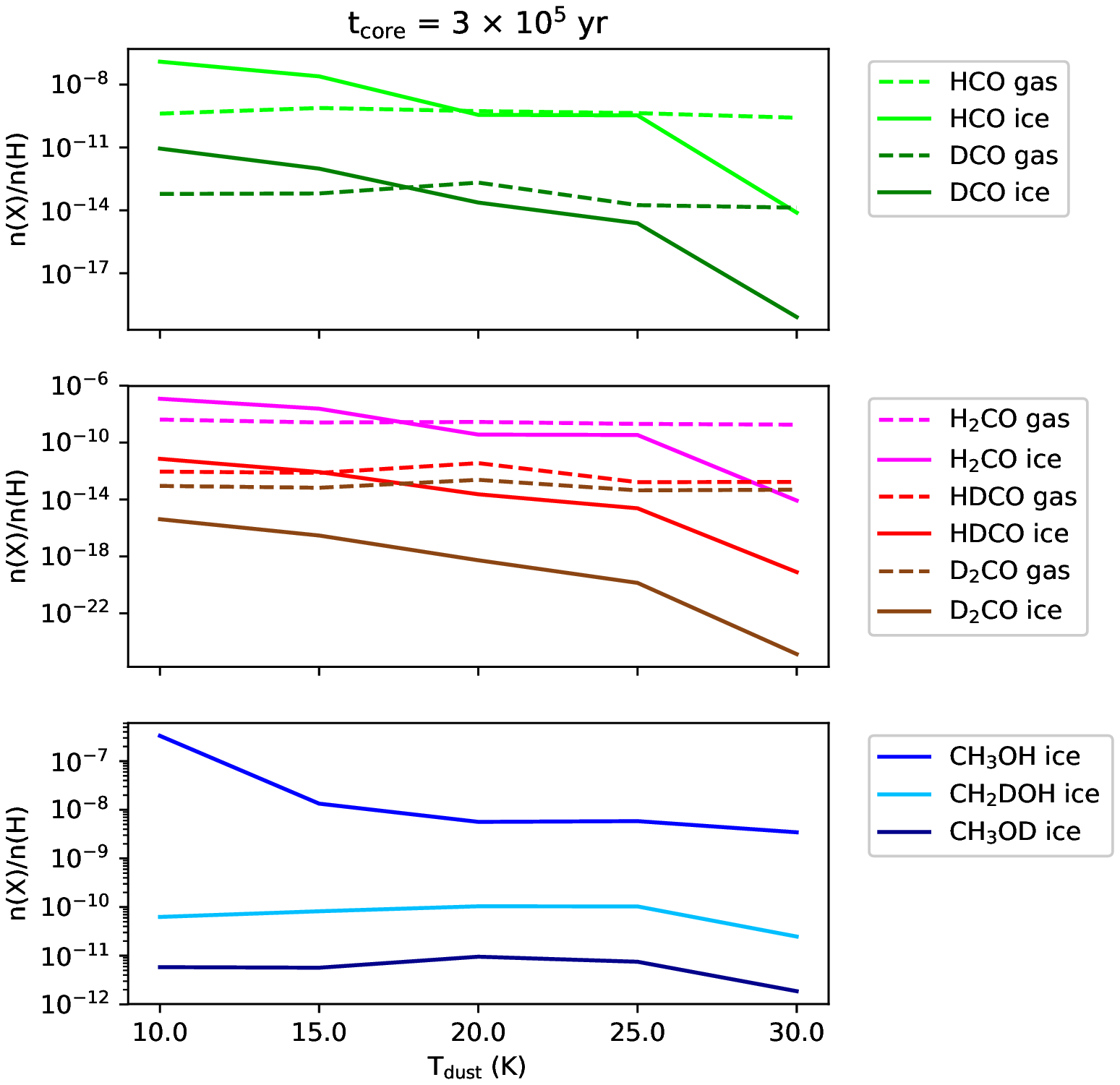}
  \caption{Abundances of key species towards the formation of deuterated methanol as obtained with Model 2B for a core age of 3~$\times$~10$^5$~yr and dust temperatures in the range of 10--30~K. Gaseous species are depicted as dashed lines, solid lines refer to grain-surface species. \ch{HCO} (light green) and \ch{DCO} (green) are shown in the upper panel. The middle panel plots \ch{H2CO} (pink), as well as \ch{HDCO} (red), and \ch{D2CO} (brown). \ch{CH3OH} (blue) and its mono-deuterated isotopologues, \ch{CH2DOH} (light blue) and \ch{CH3OD} (dark blue), are shown in the lower panel. Gas-phase abundances of (deuterated) methanol are $\leq$~10$^{-11}$ and therefore not included in the plot.}
  \label{fgr:3e5_Model2B}
\end{figure}

Grain-surface abundances of HCO, \ch{H2CO}, \ch{CH3OH}, and the deuterated isotopologues calculated with models 1, 2A, and 2B are discussed in this section for the fiducial model. Plots for Model 1, 2A, and 2B are shown in Figs. \ref{fgr:3e5_Model1}--\ref{fgr:3e5_Model2B}, and the corresponding values for the abundances of the ices are given in Tables \ref{tbl:methabunM1}--\ref{tbl:methabunM2B}. 
\newline
\noindent All considered molecules show the highest abundances if the temperatures are assumed to be low for the lifetime of the core, as grain-surface formation is most efficient at the coldest temperatures. This is because the hydrogenation and deuteration pathways on the ice rely on a source of surface H and D, which is reduced as temperatures increase. Moreover, at cold temperatures CO is depleted from the gas phase and readily available for hydrogenation and deuteration processes to start the formation pathway towards methanol \cite{Caselli12}. 

\begin{table}
    \caption{Grain-surface abundances of the species depicted in Fig. \ref{fgr:3e5_Model1}$^{\rm a}$ for t$_{\rm core}$ = 3 $\times$~10$^{5}$ yr.}
    \label{tbl:methabunM1}
    \begin{tabular}{llllll}
    \hline 
     Species    & 10 K & 15 K & 20 K & 25 K & 30K \\
     \hline 
    HCO & 1.1~$\times$~10$^{-8}$ & 1.1~$\times$~10$^{-9}$ & 2.4~$\times$~10$^{-12}$ & 1.6~$\times$~10$^{-13}$ & 3.8~$\times$~10$^{-16}$ \\
     DCO & 4.8~$\times$~10$^{-16}$ & 9.9~$\times$~10$^{-17}$ & 5.6~$\times$~10$^{-16}$ & 7.0~$\times$~10$^{-18}$ & 1.4~$\times$~10$^{-20}$ \\
     \ch{H2CO} & 1.7~$\times$~10$^{-6}$ & 1.8~$\times$~10$^{-7}$ & 3.0~$\times$~10$^{-9}$ & 3.1~$\times$~10$^{-9}$ & 2.1~$\times$~10$^{-12}$ \\
     HDCO & 2.8~$\times$~10$^{-10}$ & 3.5~$\times$~10$^{-12}$ & 4.3~$\times$~10$^{-12}$ & 1.6~$\times$~10$^{-12}$ & 1.2~$\times$~10$^{-16}$ \\
     \ch{CH3OH} & 3.0~$\times$~10$^{-5}$ & 2.1~$\times$~10$^{-7}$ & 4.3~$\times$~10$^{-8}$ & 5.2~$\times$~10$^{-8}$ & 3.0~$\times$~10$^{-8}$ \\
     \ch{CH2DOH} & 3.6~$\times$~10$^{-8}$ & 2.1~$\times$~10$^{-10}$ & 2.4~$\times$~10$^{-10}$ & 2.7~$\times$~10$^{-10}$ & 8.8~$\times$~10$^{-11}$ \\
     \ch{CH3OD} & 3.5~$\times$~10$^{-8}$ & 4.5~$\times$~10$^{-11}$ & 5.8~$\times$~10$^{-11}$ & 5.9~$\times$~10$^{-11}$ & 1.5~$\times$~10$^{-11}$ \\
    \hline 
    \end{tabular}
    \begin{tablenotes}
    \item[*] $^{\rm a}$ The shown abundances were calculated with Model 1 and the fiducial physical model for a core age of 3~$\times$~10$^5$ yr. All abundances are given relative to n$_{\rm H}$.
    \end{tablenotes}
\end{table}
\noindent For Model 1, DCO production is essentially negligible, especially compared to the obtained values of HCO, since its reaction rate proceeds slower by four orders of magnitude than the reaction that forms HCO. This is significantly lower than the values for the KIE derived from experiments and theoretical calculations, which are suggested to lie in the range of 0.004--0.08 \cite{Nagaoka07,Andersson11}. 
\newline 
\noindent The temperature dependence of the \ch{H2CO} abundance is mirrored by HDCO, but the abundance of HDCO is consistently 3--4 orders of magnitude below \ch{H2CO}. This is because D addition to HCO proceeds significantly slower than H addition to HCO. 
\noindent The main methanol isotopologue is more abundant by at least three orders of magnitude than mono-deuterated methanol. At 10~K, the produced amount of \ch{CH2DOH} and \ch{CH3OD} is more or less equal. For all other considered dust temperatures, \ch{CH2DOH} forms more abundantly than \ch{CH3OD}. The latter forms exclusively via D addition to \ch{CH3O} in the fiducial model, which competes with \ch{CH3OH} formation. 

\begin{table}
    \caption{Grain-surface abundances of the species depicted in Fig. \ref{fgr:3e5_Model2A}$^{\rm a}$ for t$_{\rm core}$ = 3 $\times$~10$^{5}$ yr.}
    \label{tbl:methabunM2A}
    \begin{tabular}{llllll}
    \hline 
    Species    & 10 K & 15 K & 20 K & 25 K & 30K \\
    \hline 
    HCO & 1.2~$\times$~10$^{-7}$ & 2.5~$\times$~10$^{-8}$ & 3.7~$\times$~10$^{-10}$ & 3.5~$\times$~10$^{-10}$ & 7.8~$\times$~10$^{-15}$ \\
    DCO & 9.3~$\times$~10$^{-12}$ & 9.8~$\times$~10$^{-13}$ & 2.4~$\times$~10$^{-14}$ & 2.6~$\times$~10$^{-15}$ & 8.2~$\times$~10$^{-20}$ \\
    \ch{H2CO} & 1.2~$\times$~10$^{-7}$ & 2.5~$\times$~10$^{-8}$ & 3.8~$\times$~10$^{-10}$ & 3.5~$\times$~10$^{-10}$ & 8.5~$\times$~10$^{-15}$ \\
    HDCO & 7.7~$\times$~10$^{-12}$ & 8.6~$\times$~10$^{-13}$ & 2.4~$\times$~10$^{-14}$ & 2.6~$\times$~10$^{-15}$ & 7.9~$\times$~10$^{-20}$ \\
    \ch{D2CO} & 4.6~$\times$~10$^{-16}$ & 2.9~$\times$~10$^{-17}$ & 5.5~$\times$~10$^{-19}$ & 1.5~$\times$~10$^{-20}$ & 1.4~$\times$~10$^{-25}$ \\
    \ch{CH3OH} & 2.4~$\times$~10$^{-7}$ & 8.9~$\times$~10$^{-9}$ & 5.6~$\times$~10$^{-9}$ & 5.8~$\times$~10$^{-9}$ & 3.4~$\times$~10$^{-9}$ \\
    \ch{CH2DOH} & 7.1~$\times$~10$^{-11}$ & 8.4~$\times$~10$^{-11}$ & 9.2~$\times$~10$^{-11}$ & 9.6~$\times$~10$^{-11}$ & 2.5~$\times$~10$^{-11}$ \\
    \ch{CH3OD} & 4.5~$\times$~10$^{-12}$ & 7.0~$\times$~10$^{-12}$ & 1.1~$\times$~10$^{-11}$ & 9.4~$\times$~10$^{-12}$ & 1.9~$\times$~10$^{-12}$ \\
    \ch{CHD2OH} & 4.5~$\times$~10$^{-12}$ & 1.2~$\times$~10$^{-12}$ & 1.6~$\times$~10$^{-12}$ & 2.0~$\times$~10$^{-12}$ & 7.2~$\times$~10$^{-13}$ \\
    \ch{CH2DOD} & 3.9~$\times$~10$^{-15}$ & 7.9~$\times$~10$^{-14}$ & 1.1~$\times$~10$^{-13}$ & 1.3~$\times$~10$^{-15}$ & 4.9~$\times$~10$^{-15}$ \\
    \ch{CHD2OD} & 7.3~$\times$~10$^{-17}$ & 7.9~$\times$~10$^{-16}$ & 1.5~$\times$~10$^{-15}$ & 1.3~$\times$~10$^{-15}$ & 9.3~$\times$~10$^{-17}$ \\
    \ch{CD3OH} & 3.0~$\times$~10$^{-13}$ & 3.1~$\times$~10$^{-14}$ & 4.9~$\times$~10$^{-14}$ & 6.6~$\times$~10$^{-14}$ & 2.2~$\times$~10$^{-14}$ \\
    \ch{CD3OD} & 4.7~$\times$~10$^{-18}$ & 1.7~$\times$~10$^{-17}$ & 4.2~$\times$~10$^{-17}$ & 3.9~$\times$~10$^{-17}$ & 2.8~$\times$~10$^{-18}$ \\
    \hline 
    \end{tabular}
    \begin{tablenotes}
    \item[*] $^{\rm a}$ The shown abundances were calculated with Model 2A and the fiducial physical model for a core age of 3~$\times$~10$^5$ yr. All abundances are given relative to n$_{\rm H}$.
    \end{tablenotes}
\end{table}
\noindent The implementation of the abstraction reaction \ch{H2CO} + H $\longrightarrow$ HCO + \ch{H2} (Table \ref{tbl:Hidakascheme}) leads to an increase of HCO in the ice by one order of magnitude in Model 2A compared to Model 1 (Fig. \ref{fgr:3e5_Model2A}). Consequently, the abundances of \ch{H2CO} are found to be lower than in Model 1.  Moreover, setting the KIE to 0.1 now forms non-negligible amounts of DCO for the lowest considered temperatures at the expense of HDCO. 
\noindent Model 2A produces less methanol than Model 1, because (1) less \ch{H2CO} is available for hydrogenation to \ch{CH2OH} and \ch{CH3O}, and (2) \ch{CH3OH} is efficiently destroyed by H abstraction occurring in the methyl group. This is in accordance with experiments that have shown that \ch{CH3OH} and \ch{H2CO} can be dehydrogenated, the products are therefore made available again for hydrogenation \cite{Chuang16} and deuteration. Abstraction of \ch{CH3OH} on the methyl side is energetically more favorable \cite{Kerkeni04}, thereby preferentially forming \ch{CH2OH}. H or D addition to this intermediate is barrierless, leading equally to either \ch{CH3OH} reformation, or the start of successive deuteration in the methyl group. This is evident in Table \ref{tbl:methabunM2A}, which shows that methyl-mono-, di-, and tri- deuterated methanol are 1--2 orders of magnitude more abundant than their counterparts with D in the hydroxy group. This holds across the entire studied temperature range.
\newline 
\noindent \ch{CH3OD} formation heavily relies on the availability of \ch{CH3O}, which is readily consumed to form \ch{CH3OH}. In the fiducial model, H addition to \ch{CH2OD} also adds significantly to the abundance of \ch{CH3OD}. However, \ch{CH2OD} either forms via D addition to \ch{H2CO}, which is slow, or directly from H abstraction of \ch{CH3OD}. Consequently, the low abundances of \ch{CH3OD} are expected within this reaction scheme. These results emphasize that the deuteration of the methyl group in methanol proceeds efficiently via H abstraction followed by D addition even within the framework of a large chemical network. 
\newline 
\noindent Many intermediate species that form multiply deuterated methanol are exclusive reactants in the deuteration scheme of methanol. Therefore, the reported values of \ch{CH2DOD}, \ch{CHD2OD}, and \ch{CD3OD} should be taken with caution, because these values may be overestimated. Reactions that reprocess these species outside of the scheme presented in Table \ref{tbl:Hidakascheme} are lacking (Section 3.3). This is also true for \ch{CD3OH}.

\begin{table}
    \caption{Grain-surface abundances of the species depicted in Fig. \ref{fgr:3e5_Model2B}$^{\rm a}$ for t$_{\rm core}$ = 3 $\times$~10$^{5}$ yr.} 
    \label{tbl:methabunM2B} 
    \begin{tabular}{llllll}
    \hline 
    Species    & 10 K & 15 K & 20 K & 25 K & 30K \\
    \hline 
    HCO & 1.2~$\times$~10$^{-7}$ & 2.4~$\times$~10$^{-8}$ & 3.6~$\times$~10$^{-10}$ & 3.4~$\times$~10$^{-10}$ & 6.7~$\times$~10$^{-16}$ \\
    DCO & 8.8~$\times$~10$^{-12}$ & 9.7~$\times$~10$^{-13}$ & 2.3~$\times$~10$^{-14}$ & 2.4~$\times$~10$^{-15}$ & 4.3~$\times$~10$^{-19}$ \\
    \ch{H2CO} & 1.2~$\times$~10$^{-7}$ & 2.4~$\times$~10$^{-8}$ & 3.7~$\times$~10$^{-10}$ & 3.4~$\times$~10$^{-10}$ & 4.2~$\times$~10$^{-15}$ \\
    HDCO & 7.3~$\times$~10$^{-12}$ & 8.6~$\times$~10$^{-13}$ & 2.3~$\times$~10$^{-14}$ & 2.4~$\times$~10$^{-15}$ & 2.3~$\times$~10$^{-18}$ \\
    \ch{D2CO} & 4.1~$\times$~10$^{-16}$ & 2.9~$\times$~10$^{-17}$ & 5.4~$\times$~10$^{-19}$ & 1.4~$\times$~10$^{-20}$ & 2.1~$\times$~10$^{-22}$ \\
    \ch{CH3OH} & 3.3~$\times$~10$^{-7}$ & 1.3~$\times$~10$^{-8}$ & 5.6~$\times$~10$^{-9}$ & 5.8~$\times$~10$^{-9}$ & 2.8~$\times$~10$^{-9}$ \\
    \ch{CH2DOH} & 6.3~$\times$~10$^{-11}$ & 8.2~$\times$~10$^{-11}$ & 1.0~$\times$~10$^{-10}$ & 1.0~$\times$~10$^{-10}$ & 2.6~$\times$~10$^{-11}$ \\
    \ch{CH3OD} & 5.8~$\times$~10$^{-12}$ & 5.7~$\times$~10$^{-12}$ & 9.5~$\times$~10$^{-12}$ & 7.5~$\times$~10$^{-12}$ & 1.6~$\times$~10$^{-12}$ \\
    \ch{CHD2OH} & 4.5~$\times$~10$^{-12}$ & 1.2~$\times$~10$^{-12}$ & 1.9~$\times$~10$^{-12}$ & 2.1~$\times$~10$^{-12}$ & 6.4~$\times$~10$^{-13}$ \\
    \ch{CH2DOD} & 4.1~$\times$~10$^{-15}$ & 6.0~$\times$~10$^{-14}$ & 9.9~$\times$~10$^{-14}$ & 7.6~$\times$~10$^{-14}$ & 6.7~$\times$~10$^{-15}$ \\
    \ch{CHD2OD} & 1.8~$\times$~10$^{-17}$ & 4.7~$\times$~10$^{-16}$ & 1.2~$\times$~10$^{-15}$ & 8.9~$\times$~10$^{-16}$ & 9.3~$\times$~10$^{-17}$ \\
    \ch{CD3OH} & 3.0~$\times$~10$^{-13}$ & 3.1~$\times$~10$^{-14}$ & 5.2~$\times$~10$^{-14}$ & 6.8~$\times$~10$^{-14}$ & 1.9~$\times$~10$^{-14}$ \\
    \ch{CD3OD} & 1.0~$\times$~10$^{-18}$ & 8.7~$\times$~10$^{-18}$ & 2.9~$\times$~10$^{-17}$ & 2.3~$\times$~10$^{-17}$ & 2.7~$\times$~10$^{-18}$ \\
    \hline 
    \end{tabular}
    \begin{tablenotes}
    \item[*] $^{\rm a}$The shown abundances were calculated with Model 2B and the fiducial physical model for a core age of 3~$\times$~10$^5$ yr. All abundances are given relative to n$_{\rm H}$.
    \end{tablenotes}
\end{table}
\FloatBarrier
\noindent HCO, DCO, \ch{H2CO}, HDCO, and \ch{D2CO} abundances of Model 2B agree with Model 2A across the considered dust temperature grid of 10--30~K. Increasing the reaction rates of the abstraction scheme leads to a slightly higher \ch{CH3OH} abundances in Model 2B at 10~K. As a consequence of the slightly more efficient formation of \ch{CH3OH}, the abundances of all deuterated isotoplogues are marginally reduced compared to Model 2A. Overall, increasing the relative rates does not impact the abundances of (deuterated) methanol significantly.
\subsection{\ch{CH2DOH}/\ch{CH3OD} Ratio}
\begin{figure}
    \includegraphics[width=1.0\textwidth]{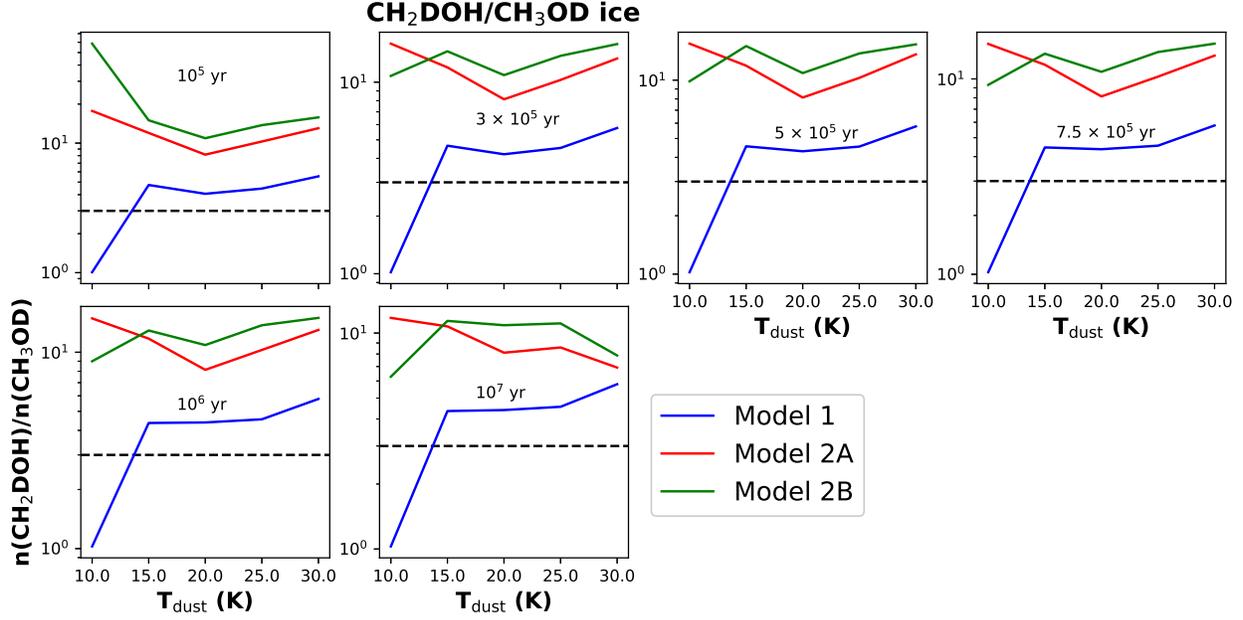}
    \caption{The grain-surface \ch{CH2DOH}/\ch{CH3OD} ratio calculated with the fiducial physical model for core ages from 10$^5$ to 10$^7$ yr. Model 1 is depicted in blue, Model 2A in red, and Model 2B in green. The black dashed line indicates the statistically expected value of 3.}
    \label{fgr:CH2DOH_CH3OD}
\end{figure}

The calculated solid \ch{CH2DOH}/\ch{CH3OD} ratios for the full physical grid can be found in Tables \ref{tbl:D1methratioM1} for Model 1, \ref{tbl:D1methratioM2A} for Model 2A, and \ref{tbl:D1methratioM2B} for Model 2B. As already discussed in the Introduction, the statistically expected value for the \ch{CH2DOH}/\ch{CH3OD} ratio is 3 \cite{Charnley97}. At the temperature most representative of a dense, prestellar core, 10~K, the \ch{CH2DOH}/\ch{CH3OD} ratio is found to be close to unity with Model 1 for core ages between 10$^5$--10$^7$ yr. Warmer dust temperatures lead to a ratio between four and six. \ch{CH3O}, the direct precursor of \ch{CH3OD}, and \ch{CH2DO}, the direct precursor of \ch{CH2DOH}, are also available for other reactions in this network. These reactions, namely the formation of (deuterated) methyl formate and dimethyl ether, start to become more efficient at warmer temperatures. Therefore, the decrease in \ch{CH3OD} is more prominent than for \ch{CH2DOH}. 
\newline 
\noindent Inclusion of the abstraction scheme leads to significantly higher \ch{CH2DOH}/\ch{CH3OD} ratios. Depending on the chosen chemical network (2A or 2B) and the initial dust temperature, the values range between 8 and 18 for the fiducial physical model (t$_{\rm core}$~=~ 3~$\times$~10$^5$~yr). At the lowest considered dust temperature of 10~K, Model 2A results in a ratio of $\sim$~16, while it is a factor of 1.5 lower in Model 2B, because Model 2A produces more \ch{CH2DOH} and less \ch{CH3OD} than Model 2B. Warmer dust temperatures (15--30~K) result in \ch{CH2DOH}/\ch{CH3OD} ratios of 8--16; and Model 2A and 2B do not differ by more than a factor of 1.5. Overall, lukewarm temperatures (15--25~K) yield higher ratios with the boosted abstraction scheme of Model 2B (i.e., inverse of what is seen at 10K). At these temperatures, \ch{CH3OD} formation is suppressed compared to \ch{CH2DOH} (as in Model 1). Boosted abstraction reaction rates in Model 2B exacerbate this. As can be seen in Fig. \ref{fgr:CH2DOH_CH3OD}, the \ch{CH2DOH}/\ch{CH3OD} ratio is mostly consistent for all calculated temperatures and core ages in the range of 10$^5$--10$^7$~yr (Tables \ref{tbl:D1methratioM2A}, \ref{tbl:D1methratioM2B}). These core ages are common choices for modelling the prestellar phase in physicochemical models \cite{Kulterer20}. 
\newline 
\noindent The boosted abstraction scheme of Model 2B leads to a very high \ch{CH2DOH}/\ch{CH3OD} ratio at 10~K and the youngest considered core age of 10$^5$~yr ($\sim$ 60 compared to 11 at 3~$\times$~10$^5$~yr). This indicates that a short, cold, prestellar core stage can lead to a very high \ch{CH2DOH}/\ch{CH3OD} ratio and that this value can be considered an upper limit. 

\begin{table}
    \caption{Solid \ch{CH2DOH}/\ch{CH3OD} ratio calculated with Model 1$^{\rm a}$.}
    \label{tbl:D1methratioM1}
    \begin{tabular}{llllll}
    \hline
     t$_{\rm core}$ (yr)   & 10 K & 15 K & 20 K & 25 K & 30K \\
     \hline 
     10$^5$ & 1.01 & 4.75 & 4.06 & 4.46 & 5.54 \\
     3 $\times$ 10$^5$ & 1.02 & 4.66 & 4.20 & 4.53 & 5.76 \\
     5 $\times$ 10$^5$ & 1.02 & 4.56 & 4.30 & 4.54 & 5.77 \\
     7.5 $\times$ 10$^5$ & 1.03 & 4.46 & 4.37 & 4.55 & 5.78 \\
     10$^6$ & 1.03 & 4.36 & 4.39 & 4.55 & 5.78 \\
     10$^7$ & 1.03 & 4.35 & 4.40 & 4.56 & 5.80 \\
    \hline 
    \end{tabular}
    \begin{tablenotes}
    \item[*] $^{\rm a}$ over the full grid of T$_{\rm dust}$ and t$_{\rm core}$ = 10$^5$ -- 10$^7$ yr in the ice 
    \end{tablenotes}
\end{table}

\begin{table}
    \caption{Solid \ch{CH2DOH}/\ch{CH3OD} ratio calculated with Model 2A$^{\rm a}$.}
    \label{tbl:D1methratioM2A}
    \begin{tabular}{llllll}
    \hline 
     t$_{\rm core}$ (yr)   & 10 K & 15 K & 20 K & 25 K & 30K \\
     \hline 
     10$^5$ & 17.71 & 11.98 & 8.15 & 10.29 & 13.03 \\
     3 $\times$ 10$^5$ & 15.88 & 11.94 & 8.14 & 10.27 & 13.29 \\
     5 $\times$ 10$^5$ & 15.42 & 11.87 & 8.13 & 10.27 & 13.58 \\
     7.5 $\times$ 10$^5$ & 15.10 & 11.82 & 8.13 & 10.27 & 13.16 \\
     10$^6$ & 14.87 & 11.73 & 8.14 & 10.27 & 12.99 \\
     10$^7$ & 11.76 & 10.74 & 8.11 & 8.58 & 6.92 \\
    \hline 
    \end{tabular}
    \begin{tablenotes}
    \item[*]$^{\rm a}$ over the full grid of T$_{\rm dust}$ and t$_{\rm core}$ = 10$^5$ -- 10$^7$ yr in the ice 
    \end{tablenotes}
\end{table}

\begin{table}
    \caption{Solid \ch{CH2DOH}/\ch{CH3OD} ratio calculated with Model 2B$^{\rm a}$.}
    \label{tbl:D1methratioM2B}
    \begin{tabular}{llllll}
    \hline 
     t$_{\rm core}$ (yr)   & 10 K & 15 K & 20 K & 25 K & 30K \\
     \hline 
     10$^5$ & 58.56 & 14.99 & 10.91 & 13.76 & 15.80 \\
     3 $\times$ 10$^5$ & 10.78 & 14.49 & 10.88 & 13.72 & 15.80 \\
     5 $\times$ 10$^5$ & 9.86 & 14.99 & 10.87 & 13.73 & 15.29 \\
     7.5 $\times$ 10$^5$ & 9.32 & 13.46 & 10.87 & 13.72 & 15.15 \\
     10$^6$ & 8.99 & 12.90 & 10.88 & 13.72 & 14.96 \\
     10$^7$ & 6.28 & 11.39 & 10.87 & 11.09 & 7.88 \\
    \hline 
    \end{tabular}
    \begin{tablenotes}
    \item[*] $^{\rm a}$ over the full grid of T$_{\rm dust}$ and t$_{\rm core}$ = 10$^5$ -- 10$^7$ yr in the ice 
    \end{tablenotes}
\end{table}

\FloatBarrier

\subsubsection{Parameter Space Exploration at 10~K}
Independent of the choice of the chemical model, the results over the considered range of physical parameters have shown that dust temperature is the most influential factor on the \ch{CH2DOH}/\ch{CH3OD} ratio. However, average dust temperatures in the inner parts of prestellar cores are typically $\leq$~10~K and typical gas densities are found to be in the range of 10$^4$--10$^6$~cm$^{-3}$ \cite{Benson89,Bergin07}. Moreover, previous modelling work has suggested that an activation barrier of $<$~1~500~K for the reaction CO + H $\longrightarrow$ HCO is needed to match models to observations \cite{Taquet12}. In this section, a range of gas densities and activation barriers E$_{\rm A(CO+H)}$ is explored for a fixed dust temperature of 10 K (Table 2).
\newline
\noindent Calculated \ch{CH2DOH}/\ch{CH3OD} ratios with Model 1 are solely shown in the Supporting Information (Table S1, Fig. S1). Independent of the core age and E$_{\rm A}$, the ratio remains between 1 and 2 in Model 1, if the gas density is $\leq$~10$^5$~cm$^{-3}$. For higher gas densities and core ages $>$~10$^5$~yr, the ratio can go up to $\sim$~5, with higher values corresponding to the lower end of the considered E$_{\rm A}$ range.
\newline 
\noindent Higher gas densities and lower activation barriers for the CO + H $\longrightarrow$ HCO reaction increase the rate of this reaction, and therefore all the relative rates in the abstraction scheme in Model 2A. The considered grid of gas densities and E$_{\rm A}$ is used to investigate variations of the \ch{CH2DOH}/\ch{CH3OD} ratio with the newly constructed Models 2A$\_$\ch{CX2OX} and 2A$\_$Song. Combined with the aforementioned parameter space explored by the theoretical modelling work of T12 \cite{Taquet12}, these models also test the reaction \ch{H2CO} + H based on the E$_{\rm A}$ for the formation of \ch{CH2OH}. Adopted values are taken from the \textsc{kida} database \cite{Wakelam12}, and quantum chemical calculations \cite{Song17}. Those values were published after Hidaka and colleagues conducted their experiments \cite{Nagaoka07,Hidaka09}. 
\begin{figure}
    \centering
    \includegraphics[width=0.8\textwidth]{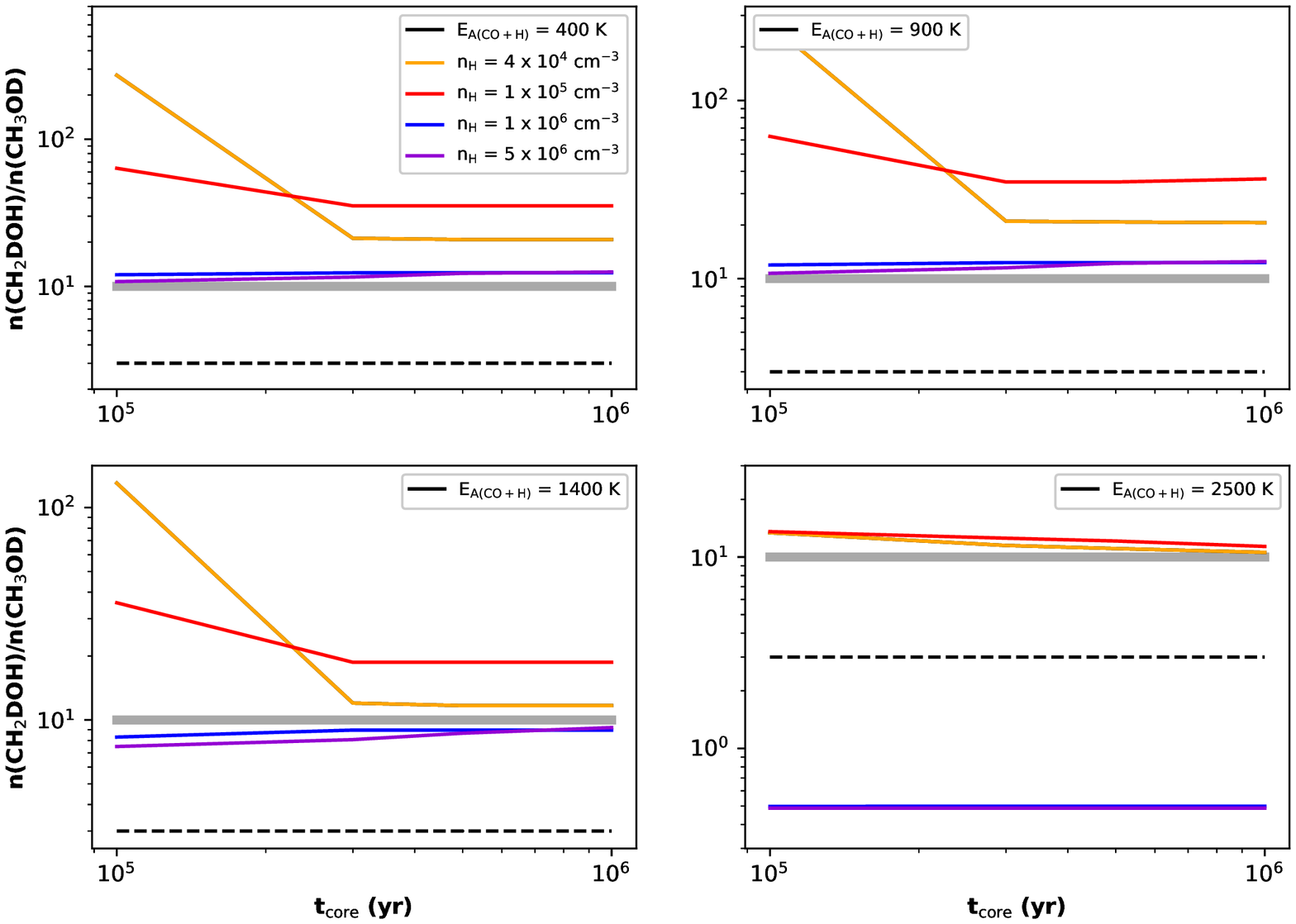}
    \caption{Calculated \ch{CH2DOH}/\ch{CH3OD} ratios over a range of gas densities and activation barriers for the reaction CO + H $\longrightarrow$ HCO at T$_{\rm dust}$ = 10 K with Model 2A$\_$\ch{CX2OX} for core ages ranging from 10$^5$--10$^6$ yr. The dashed black line refers to the statistically expected value of 3, the solid grey line refers to the lower limit of this ratio determined from observations of the prestellar core L1544 \cite{Bizzocchi14}.}
    \label{fgr:GCH2DOH_GCH3OD_CX2OX}
\end{figure}
\noindent Results for Model 2A$\_$\ch{CX2OX} at 10~K for core ages between 10$^5$--10$^6$~yr are shown in Fig. \ref{fgr:GCH2DOH_GCH3OD_CX2OX}. The values for the \ch{CH2DOH}/\ch{CH3OD} ratio are detailed in Table S2. If E$_{\rm A}$ is below 1~400~K, the \ch{CH2DOH}/\ch{CH3OD} ratio is found to be as high as $\sim$~200 for low gas densities. While the \ch{CH2DOH} abundances remain comparable to those in Model 2A,  \ch{CH3OD} is efficiently destroyed by the high relative rates of the abstraction scheme, and its precursor \ch{CH3O} is more likely to react with H to form methanol than to react with D to form \ch{CH3OD}. On the other hand, gas densities $>$~10$^6$~cm$^{-3}$, which are expected to be found in the inner regions of prestellar cores on the verge of collapse, give \ch{CH2DOH}/\ch{CH3OD} ratios of $\sim$~7--12. The picture turns if the gas density grid is explored with the E$_{\rm A}$ value of the fiducial model, 2~500~K. Here, high gas densities result in a \ch{CH2DOH}/\ch{CH3OD} ratio of $\sim$~0.5, and low gas densities results in ratios of $\sim$~11--16. This highlights the importance of not only exploring an expanded grid of physical parameters, but also the full range of possible activation barriers in key reactions that are used by the community.
\newline 
\begin{figure}
    \centering
    \includegraphics[width=0.8\textwidth]{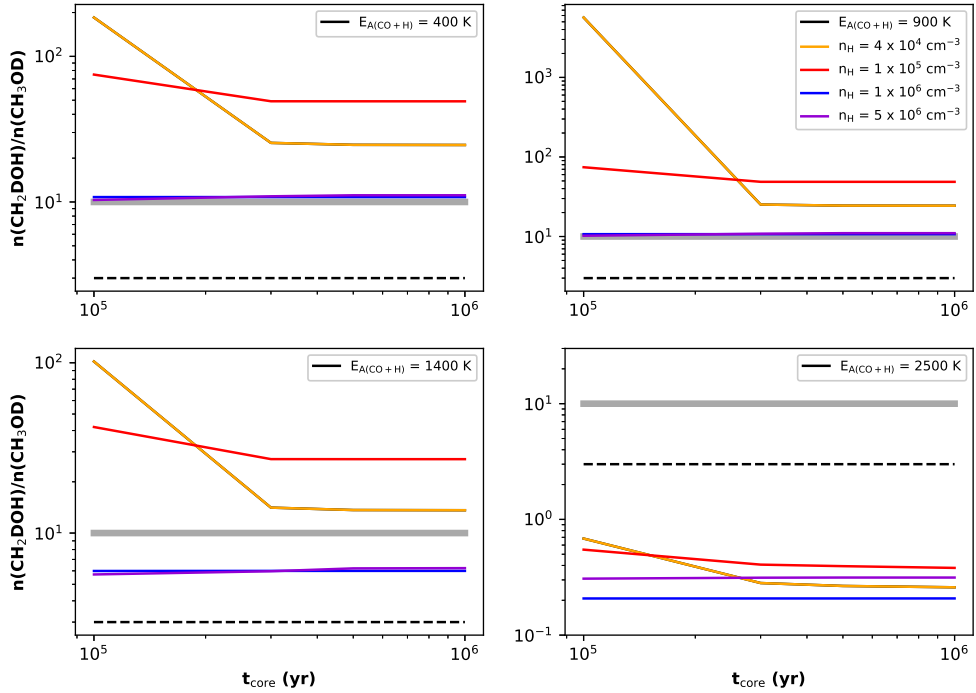}
    \caption{Calculated \ch{CH2DOH}/\ch{CH3OD} ratios over a range of gas densities and activation barriers for the reaction CO + H $\longrightarrow$ HCO at T$_{\rm dust}$ = 10 K with Model 2A$\_$Song for core ages ranging from 10$^5$--10$^6$ yr. The dashed black line refers to the statistically expected value of 3, the solid grey line refers to the lower limit of this ratio determined from observations of the prestellar core L1544 \cite{Bizzocchi14}.}
    \label{fgr:GCH2DOH_GCH3OD_Song}
\end{figure}
\noindent The same range of core ages, gas densities, and activation barriers is also explored with Model 2A$\_$Song. Results are depicted in Fig. \ref{fgr:GCH2DOH_GCH3OD_Song}, the values are tabulated in Table S3. Similar to Model 2A$\_$\ch{CX2OX}, low gas densities and low E$_{\rm A}$ result in a high \ch{CH2DOH}/\ch{CH3OD} ratio (up to $\sim$~100), especially at core ages of 10$^5$~yr. High density models that are calculated with E$_{\rm A}$~$\leq$~1~400~K result in ratios of $\sim$~6--11, consistent with findings of Model 2A$\_$\ch{CX2OX}. The most distinct difference is found for calculations carried out for E$_{\rm A}$~=~2~500~K for the considered gas density grid, which leads to \ch{CH2DOH}/\ch{CH3OD} ratios below unity. This can be explained by the suppressed dissociation reactions  of \ch{H2CO}, HDCO, and \ch{D2CO} in Model 2A$\_$Song, which lead to less formation of HCO and DCO when compared to Model 2A$\_$\ch{CX2OX}. DCO is the main precursor of HDCO, which in turn reacts with 2H and forms \ch{CH2DO}, and subsequently \ch{CH2DOH}. Therefore, lower \ch{CH2DOH} abundances are computed with this model. At lower E$_{\rm A}$, two other formation pathways can compensate for the lack of HDCO formation. In what is denoted as S3, Eq. \ref{eqn:S3}, a deuteration sequence of \ch{H2CO} with HCO and DCO as intermediates, is the main product channel of HDCO formation, amplified for the lower end of the gas density grid. Moreover, at low E$_{\rm A}$, CO + D is fast and produces enough DCO to form \ch{CH2DOH} comparable with Model 2A$\_$\ch{CX2OX}. This emphasizes that it is crucial to carefully evaluate the formation of deuterated methanol via abstraction and substitution reactions and the reactions leading to the formation of the precursor molecules. 
\FloatBarrier 
\subsection{Network caveats}
It is well established that COMs, which are synthesized on grain surfaces, are also detected towards prestellar cores, where the temperatures are $\sim$~10~K \cite{Bacmann12,JimenezSerra16}. As they are observed in the gas phase, non-thermal mechanisms that take them off the grains must be at work. One such mechanism, which is yet to be considered in the presented models, is reactive desorption. In this process, the energy released by chemical reactions on grain surfaces is not fully absorbed by the grain; and the remaining energy causes the ejection of the newly formed product into the gas phase \cite{Pantaleone20}. Chemical models usually assume a fixed percentage for all formed species \cite{Walsh14}; the contribution of reactive desorption is often set to 1\% \cite{Garrod07,Vasyunin13}, but also sometimes as high as 10\% \cite{Walsh14}. Experiments that specifically investigate reactive desorption in the CO--\ch{H2CO}--\ch{CH3OH} chain with H addition and abstraction reactions match the range of 1--10\% assumed by chemical models \cite{Chuang18}. On the other hand, the theoretical investigation of the energy release of the CO + H reaction, the starting point of methanol formation, has shown that the formed HCO is left with an energy content that is about two times lower than its adsorption energy \cite{Pantaleone20}. Therefore, reactive desorption is inefficient for the CO + H system, at least on crystalline ice. These calculations also suggest that conducted these calculations also speculate that this behaviour is quite general for hydrogen bonds. Therefore, not including reactive desorption in the current network should not influence the grain-surface abundances of (deuterated) HCO, \ch{H2CO}, and \ch{CH3OH} discussed here. Moreover, reaction rates obtained from the non-thermal desorption mechanism that are included in the current version of the network are slow compared to the grain-surface reaction rates, and therefore, do not impact the ice abundances significantly. However, gas-phase abundances obtained with the networks presented here are likely lower limits.   
\newline 
\newline 
\noindent The abstraction scheme towards the formation of deuterated methanol is fully reliant on the reaction rate of CO + H that goes into the model. Tunneling in this reaction, and in CO + D, have been investigated via harmonic quantum transition state theory \cite{Andersson11}. This work showed that tunneling in these reactions is significant. The reaction rate for hydrogenation of CO assumed in the fiducial is about a factor of 2 higher for the gas density of 4~$\times$~10$^4$~cm$^{-3}$ and an activation barrier of 2~500~K than what these authors calculate for this reaction at 10~K and $\sim$~1.1 times higher at 20~K. This indicates that the adopted fiducial model results are computed with a reasonable reaction rate. On the other hand, the investigated range of gas densities and activation barriers for the hydrogenation of CO (Section 3.2.1) can increase the reaction rate of CO + H by up to five orders of magnitude. However, the quantum chemical calculations \cite{Andersson11} find a KIE of $\sim$~0.004 for this reaction at low temperatures (5--20~K), which is contrasting to the KIE of 0.08--0.1 obtained from the experiments that serve as the basis of the abstraction scheme \cite{Nagaoka05,Nagaoka07,Hidaka09}. The KIE derived from experiments stems not only from the tunneling reaction, but also from other contributions to surface reactions, such as the longer residence time of D on the grains due to its higher mass. If these contributions enhance the reactivity, the calculated KIE \cite{Andersson11} can still be valid, and the chosen value of 0.1 for the KIE in this work is more representative of the processes on an actual interstellar grain. 
\newline 
\noindent An underlying assumption in the laboratory experiments is that the gas that accretes onto the grain surfaces has a D/H ratio of 0.1 \cite{Nagaoka05,Nagaoka07,Watanabe06,Hidaka09}. The level of alteration of the relative reaction rates (Table \ref{tbl:Hidakascheme}) as a result of a different D/H in the accreting gas is unclear. Moreover, the experiment was conducted on ASW surfaces at 15~K. The considered surface and temperature do influence the reactions \cite{Watanabe06}; hence, if the composition of the interstellar ices differs significantly from ASW, the reaction rates may not be accurate anymore. It is also to note that this work models dust temperatures in the range of 10--30~K under the assumption that the relative rates derived at 15~K remain unchanged across this range. More experiments are needed to clarify the influence of the surface composition, the underlying D/H ratio, and the surface temperature. 
\newline 
\newline 
\noindent A caveat of the single-point approach of this model is that the calculated D/H ratios could be too low. If one would want to investigate the D/H ratio, it would be advisable to first construct a gas-phase network that forms atomic deuterium, starting with the reaction \ch{H3+} + HD $\longrightarrow$ \ch{H2D+} + \ch{H2} in a physical model that mimics an earlier evolution stage, before freeze-out of CO and its subsequent hydrogenation and deuteration become efficient. In a second step, the fiducial model could then be coupled to the grain-surface network of Models 1, 2A, and 2B, similar to what has been done in a study with a 3-phase network \cite{Taquet12}. On the other hand, the D/H ratios of water ice in Models 1, 2A, and 2B are in the range of 10$^{-4}$, which is in good agreement with models investigating the deuteration of simple molecules at the center of starless cores \cite{Sipila13}. 
\newline 
\noindent Nonetheless, the ratios of \ch{CH2DOH}/\ch{CH3OH} and \ch{CH3OD}/\ch{CH3OH} are low at 10~K if compared to observations of low-mass protostars \cite{Parise06survey}, more so for Models 2A and 2B. Other models that do not extend the complexity of the grain-surface network beyond methanol obtain a better agreement to the observed D/H values \cite{Taquet13}. In the models presented here, some of the atomic D is used to deuterate other grain-surface species. This limits the amount of D that is available for a) the formation of deuterated methanol via addition reactions, b) inducing H abstraction from the methyl group of methanol, and c) destroying the main methanol isotopologue.
\newline 
\noindent However, this work is interested in investigating the impact of a variety of parameters on the  \ch{CH2DOH}/\ch{CH3OD} ratio and its potential of being a tracer of the prestellar stage, not the D/H ratio of the different flavors of deuterated methanol. A higher availability of D in the abstraction scheme could potentially lead to a more efficient destruction of (deuterated) methanol by H abstraction from the methyl group (see reactions in Table \ref{tbl:Hidakascheme}), for e.g., \ch{CH2DOH} via the \ch{CH2DOH} + H $\longrightarrow$ CHDOH + HD reaction. Subsequent H addition will replenish the previously destroyed \ch{CH2DOH} via CHDOH + H $\longrightarrow$ \ch{CH2DOH}, while D addition will increase the deuteration degree of methyl-deuterated methanol via the \ch{CHDOH} + D $\longrightarrow$ \ch{CHD2OH} reaction. Hence, a higher D/H ratio available for surface chemistry on the grains will increase methyl-deuterated methanol formed by the network and suppress the formation of hydroxy-deuterated methanol compared to the findings of this work. Therefore, the conclusion that the abstraction scheme leads to a \ch{CH2DOH}/\ch{CH3OD} ratio greater than the statistically expected value of 3 still holds, and is probably even enhanced.
\newline 
\newline 
\noindent 
Even though the underlying chemical network in Models 2A and 2B is heavily tailored towards the formation of deuterated methanol, which made the addition of multiply deuterated isotopologues for species outside of this system necessary, the overall results of the networks are still reasonable. Comparing the obtained values for gas-phase CO with observations of CO in prestellar cores \cite{Caselli08} validates Models 1, 2A, and 2B at 10~K, and cloud ages in the range of 10$^5$--10$^6$~yr. For instance, values for the fiducial model at 10~K are in the range of 2~$\times$~10$^{-5}$, whereas observations give values between 0.3--3.0~$\times$~10$^{-5}$. Moreover, nine simple species across three different physicochemical models have been previously analysed \cite{Kulterer20}. If the abundances of all deuterated isotopologues and the main isotopologue are summed for this set of species, Models 1, 2A, and 2B are within a factor of 2 (excluding \ch{H2CO} and \ch{CH3OH}, which are 1--2 orders of magnitude lower). The largest differences between the comparative work of three different models \cite{Kulterer20} and Models 1, 2A, and 2B are seen for solid \ch{H2S} and \ch{CH4} (up to two orders of magnitude). However, they are not due to the inclusion of the abstraction scheme, because the abundances of solid \ch{H2S} and \ch{CH4} are consistent between Models 1, 2A and 2B.

\FloatBarrier

\section{Discussion}
\subsection{Deficiency of \ch{CH3OD}}
The vast majority of the models presented in this work produce less \ch{CH3OD} compared to \ch{CH2DOH}. Only specific combinations of gas densities, core ages, and considered activation barriers (in Models 2A$\_$\ch{CX2OX} and 2A$\_$Song), or a simplified chemical model (Model 1) lead to roughly equal abundances of \ch{CH2DOH} and \ch{CH3OD}. \ch{CH3OD} formation is low in the abstraction schemes of Models 2A and 2B, especially at low temperatures. This is amplified when more recent results for the non-barrierless reactions (formation of (deuterated) HCO and \ch{CH3OH}) in the methanol system are included in the models. In the most extreme cases, \ch{CH3OD} can be as much as a factor of $\sim$~200 less abundant than \ch{CH2DOH}.
\newline 
\noindent \ch{CH3OD} formation is also not seen in low-temperature experiments, while the formation of methyl-deuterated methanol is detected \cite{Nagaoka05} with \ch{CH3OH} as the starting point. 
\newline 
\noindent These experiments do conclusively form methanol ice at cold temperatures and indicate that hydrogenation of CO towards \ch{CH3OH} via \ch{H2CO} proceeds efficiently below 12~K in CO and CO-\ch{H2O} mixed ices \cite{Watanabe04,Fuchs09}. For temperatures in the range of 12--15~K, the formation of \ch{H2CO} and \ch{CH3OH} is still seen in the mixed ice, whereas it is hindered in pure CO ice due to the diffusion length of H. Only the upper \ch{H2CO} layers on top of the CO ice can be processed to form \ch{CH3OH} \cite{Watanabe04}. At temperatures $>$~20~K, \ch{CH3OH} formation starting from CO hydrogenation is inefficient \cite{Watanabe04,Watanabe06}. Thus, especially for the calculations carried out in the temperature range of 10--15~K, \ch{CH3OH} is readily available on the ice surface to start the successive deuteration of its methyl group (Eq. \ref{eqn:S2}).
\newline 
\noindent A seperate set of laboratory experiments at 15~K detected that H abstraction from \ch{CH3OH} and \ch{H2CO} can lead to dehydrogenation and reformation of the methoxy and formyl radicals, and CO. These radicals would then be available to form more complex species, reform methanol, or be deuterated \cite{Chuang16}. This contradicts the theoretical findings of Song and colleagues \cite{Song17} that show that the formation of the formyl radical upon H abstraction of \ch{H2CO} is suppressed. The discrepancy may stem from the theoretical calculations being carried out on an ASW surface and the experiments being executed for mixed ices of CO, \ch{H2CO}, and \ch{CH3OH}. Nevertheless, the formation of \ch{H2CO} and \ch{CH3OH} alongside the formation of species with a higher degree of complexity (with more than one C atom) is also confirmed by additional experiments \cite{Fedoseev15}. The results presented in Fedoseev et al. (2015) \cite{Fedoseev15} attribute the formation of these species to the recombination of HCO and \ch{CH3O}, which puts emphasis on the importance of the dehydrogenation of \ch{H2CO} and \ch{CH3OH} \cite{Chuang16}. Additionally, it has been shown that strong competition between H-addition and H-abstraction takes place for the H + \ch{H2CO} reaction and the reactive desorption of \ch{H2CO} \cite{Minissale16}. 
\newline 
\noindent Another formation mechanism that has so far only been investigated experimentally is methanol formation via oxygen insertion in ices below the oxygen desorption temperature of $\sim$~25~K \cite{Bergner17}. This study shows that upon the dissociation of \ch{O2} (\ch{O2} $\longrightarrow$ O + O), $\sim$~65\% of oxygen insertion into \ch{CH4} ice leads to \ch{CH3OH} (O + \ch{CH4} $\longrightarrow$ \ch{CH3OH}), while the remainder forms \ch{H2CO}. This formation pathway has not been included in models, but should definitely be explored in the future. However, analogous experiments that study oxygen insertion into mono-deuterated methane to investigate the formation of mono-deuterated methanol have not been carried out yet. 
\newline 
\noindent Including the aforementioned full abstraction sequence from \ch{CH3OH} back to CO and the oxygen insertion mechanism need to be included in future models to obtain a more complete picture of methanol formation and its deuterated isotopologues on grain surfaces.
\newline 
\newline 
\noindent Furthermore, the formation of \ch{H2CO} and \ch{CH3OH} has been studied by quantum chemical electronic structure calculations on water ice \cite{Woon02}. The influence of water ice on the barrier of the H + CO reaction was determined to be negligible. However, for reaction  H + \ch{H2CO} it enhances \ch{CH3O} production, whereas it hinders the abstraction reaction leading to \ch{H2} + HCO, again confirming the findings of Song et al. (2017) \cite{Song17}. A different substrate might change the outcome though, as the ice structure has been shown to have a huge impact on the results in experiments \cite{Watanabe06}. Pure CO ice might be the more efficient substrate for hydrogenation, but it has been argued that realistic CO ices likely contain impurities from OH groups like methanol already, which should have a similar effect on reactions as if they were occurring on ASW \cite{Song17}. 
\newline 
\noindent Experiments \cite{Nagaoka07} showed that deuteration towards \ch{CD3OH}, with \ch{CH2DOH} and \ch{CHD2OH} formation on the way, starts by a repetition of the formation of a hydroxymethyl radical by the abstraction of an H atom from methanol and D addition to the hydroxymethyl radical, not from intermediates that stem from H-D addition to (deuterated) formaldehyde. Nagaoka et al. (2007) \cite{Nagaoka07} observe abundance maxima in the obtained infrared absorption spectra for \ch{CH3OH} and \ch{CH2DOH}, therefore they get destroyed to form multiply deuterated methanol in the experiments at 15~K with an atom flux at a D/H ratio of 0.1. It is not possible to verify whether repeated abstraction and addition or direct exchange is responsible for the formation of deuterated methanol by experiments, but direct exchange has higher activation energies \cite{Blowers98}. The non-detection of deuteration in the hydroxy group in these experiments \cite{Nagaoka05} can be explained by the fact that the barrier for abstraction with both, H and D, is significantly lower on the methyl side \cite{Kerkeni04}. Also, the dissociation energy of the O-H bond is larger (\ch{CH3OH} $\longrightarrow$ \ch{CH3O} + H) than that of the C-H bond (\ch{CH3OH} $\longrightarrow$ \ch{CH2OH} + H) in \ch{CH3OH} \cite{Nagaoka07}. Finally, as a result of tunneling, H addition occurs at faster rates than D addition, therefore H addition to \ch{CH3O} is more likely than the D addition channel that would form \ch{CH3OD} \cite{Watanabe06}.
\newline 
\noindent The abstraction reaction H + \ch{CH3OH} $\longrightarrow$ \ch{H2} + \ch{CH2OH}, \ch{CH3O} has also been studied theoretically at temperatures down to 30~K for various isotopologues \cite{Goumans11methpaper} to quantify the impact of atom tunneling on the reaction rates. Extrapolating these results down to 10~K shows that abstraction via H proceeds more efficiently than abstraction via D. Moreover, abstraction from the C atom of \ch{CH3OH} is always preferred over abstraction from the O atom more strongly at low temperatures \cite{Goumans11methpaper}, which is in accordance with the findings of the previously described experiments \cite{Nagaoka05,Nagaoka07}. This reaction is not specifically included in the abstraction scheme presented by Hidaka and his colleagues \cite{Hidaka09} (Table \ref{tbl:Hidakascheme}) and not implemented in the current version of the network, because the aim of this work is to explicitly test this abstraction scheme. Not included are also the abstraction reactions from \ch{CH2OH} and \ch{CH3O} that re-form \ch{H2CO}, and the abstraction reaction from HCO that leads back to CO, which have been found to proceed in experiments \cite{Chuang16}. The impact of including these reactions will be examined in a future project.
\newline 
\noindent 
The evidence presented here should establish the importance of including abstraction reactions in any network that wants to accurately model the formation of deuterated methanol in the prestellar phase.  A simple addition model, like Model 1, is missing key formation mechanisms that should be implemented in line with an array of experimental and theoretical studies. As a consequence of the investigated scheme, \ch{CH3OD} production is much more limited than that of \ch{CH2DOH} in prestellar cores. The \ch{CH2DOH}/\ch{CH3OD} ratio does not seem to be an adequate tracer of prestellar phase conditions.
\newline 
\noindent On the other hand, the \ch{CH2DOH}/\ch{CH3OH}, \ch{CHD2OH}/\ch{CH2DOH}, \ch{CD3OH}/\ch{CHD2OH}, and \ch{CD3OH}/\ch{CH2DOH} ratios are suitable prestellar tracers. Experiments \cite{Nagaoka05,Nagaoka07,Hidaka09} demonstrate that the abstraction scheme is efficient at cold, prestellar conditions and most likely not relevant at later, warmer stages, because the diminishing availability of CO on the grains hinders the formation of methanol. Therefore, the predominant formation of methyl-deuterated methanol is expected on grain surfaces in prestellar cores. Variations in the ratio of methyl-deuterated over non-deuterated methanol are expected from core to core depending on the elemental atomic deuterium abundance, A$_{\rm D}$, the core age, the gas density, and the dust temperature, because these parameters determine the efficiency of this scheme. The idea that multiply deuterated isotopologues have the potential to constrain the D/H ratio in prestellar cores has been proposed for the \ch{D2O}/HDO ratio \cite{Furuya17}. However, \ch{CHD2OH} has not been detected in starless or prestellar cores, and only in seven protostellar systems \cite{Parise02double,Agundez19,Drozdovskaya22}. Carrying out observations of \ch{CHD2OH} in prestellar cores and in additional protostellar systems should provide insights on whether the \ch{CHD2OH}/\ch{CH2DOH} ratio is a candidate for constraining natal cloud environments.

\subsection{Comparison with prestellar core observations}
\begin{figure}
    \centering
    \includegraphics[width=0.8\textwidth]{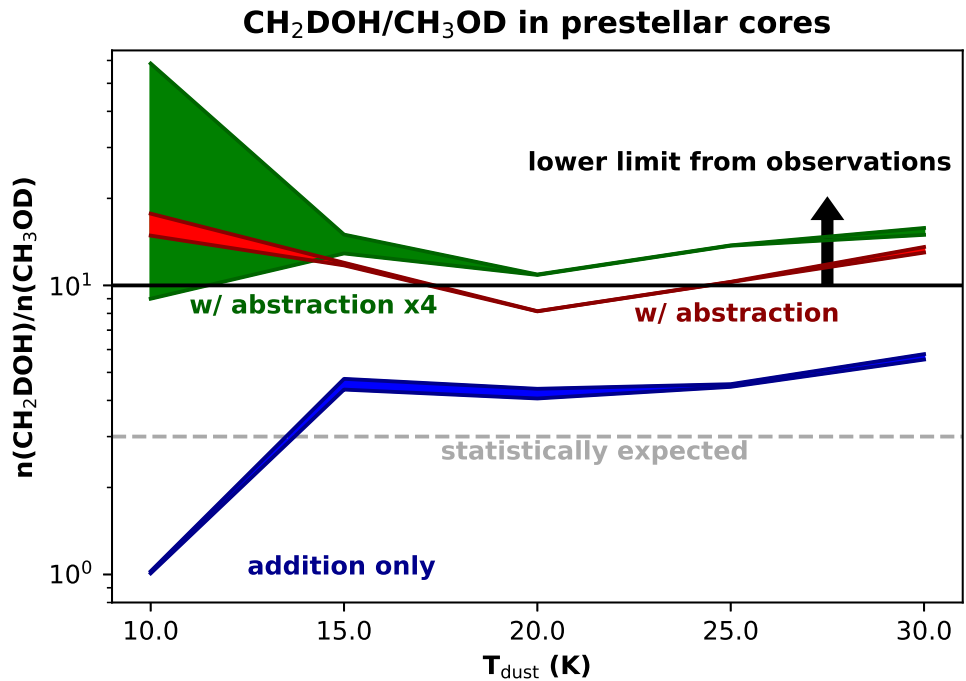}
    \caption{Calculated \ch{CH2DOH}/\ch{CH3OD} ratios compared with what is expected to be found in prestellar cores from observations. Values are depicted for a T$_{\rm dust}$ range of 10--30~K and core ages of 10$^5$--10$^6$~yr. The blue slope, denoted ``addition only'' refers to results from Model 1, the red slope (``w/ abstraction'') describes results from Model 2A, and the green slope (``w/ abstraction x4'') stems from results from Model 2B. The solid black line indicates the lower limit of the \ch{CH2DOH}/\ch{CH3OD} ratio as derived towards the dust peak of the prestellar core L1544 \cite{Bizzocchi14}. The dashed grey line highlights the statistically expected value of three.}
    \label{fgr:GCH2DOH_GCH3OD_obscomp}
\end{figure}

In this work, the \ch{CH2DOH}/\ch{CH3OD} ratio varies from 1 to 200 across the entire considered range of models. The simplistic additions-only Model 1 obtains a ratio of $\sim$~1 at 10~K for the majority of the calculated models. For some combinations of high gas densities, low E$_{\rm A}$, and warmer gas temperatures, the ratio can go up to $\sim$~6. Depending on whether Model 2A or 2B (non-boosted or boosted abstraction scheme) is considered, low temperatures lead to a ratio of $\sim$~60 or $\sim$~18, respectively. Both models result in ratios of 7--16 for warmer temperatures in the range of 15--30~K. Additional calculations exploring a range of gas densities and activation barriers at 10~K also find \ch{CH2DOH}/\ch{CH3OD} ratios predominately in the range of 6--200.  
\newline 
\noindent Gas-phase detections of mono-deuterated methanol in starless and prestellar cores are sparse. \ch{CH2DOH} has been detected in a few sources at column densities of $\sim$~4--20\% relative to \ch{CH3OH} \cite{Ambrose21,Bizzocchi14,Lattanzi20}; however, only one upper limit for \ch{CH3OD} in the well-studied prestellar core L1544 has been reported in the literature \cite{Bizzocchi14} ($<$1\% w.r.t. to methanol). 
\newline 
\noindent The lack of \ch{CH3OD} in the prestellar phase does not allow a lengthy discussion of the \ch{CH2DOH}/\ch{CH3OD} ratio. The lower limit reported in L1544 \cite{Bizzocchi14} leads to \ch{CH2DOH}/\ch{CH3OD}~$\geq$~10 from column densities derived for excitation temperatures T$_{\rm ex}$~=~5--8~K. As can be seen in Fig. \ref{fgr:GCH2DOH_GCH3OD_obscomp}, this confirms the conclusion that simple addition reactions alone (Model 1), which lead to a ratio of $\sim$~1 at 10~K across the considered core ages, are not sufficient for explaining the observed abundances. 

\noindent \ch{CH2DOH} has been detected in a number of prestellar cores in recent years with single dish telescopes. A survey of 12 starless and prestellar cores in the B10 region in the Taurus Molecular Cloud revealed \ch{CH2DOH} in 9 out of the 12 sources \cite{Ambrose21}. Additionally, \ch{CH2DOH} has been detected in the prestellar cores L183 \cite{Lattanzi20} and L1544 \cite{Bizzocchi14}. L1544 has been the target of many extensive studies \cite{Keto10,JimenezSerra16} and was the first prestellar core with a reported \ch{CH2DOH} detection \cite{Bizzocchi14}, which was derived towards the dust peak of L1544. Mapping observations have revealed a molecular differentiation in L1544 \cite{Spezzano17}; depending on whether observations were carried out towards the dust peak at the center of the core or towards the so-called methanol peak, the abundance ratio N(\ch{CH2DOH})/N(\ch{CH3OH}) is in the range of 0.06--0.13 \cite{ChaconTanarro19}. 
\newline
\noindent The numerous detections of \ch{CH2DOH}, but their lack for \ch{CH3OD} fit the narrative painted by experiments and modelling that prestellar cores are deficient in \ch{CH3OD}. It remains unclear whether the abstraction scheme implemented here is fit to reproduce the \ch{CH2DOH}/\ch{CH3OD} ratio in prestellar cores. Only Model 1 is conclusively ruled out. A firm detection of \ch{CH3OD} in a source of this type will provide a crucial reference point to shed light on this open question.
\newline 
\noindent Mono-deuterated methanol has also been detected in comet 67P/Churyumov–Gerasimenko using \textit{Rosetta}-ROSINA data \cite{Drozdovskaya21}. The ROSINA Double Focusing Mass Spectrometer cannot distinguish  \ch{CH2DOH} from \ch{CH3OD}; as a result, it is not possible to confirm whether both isotopologues are detected. In the unrealistic scenario that no \ch{CH2DOH} is present in the comet, the abundance of \ch{CH3OD} would be 5.5\% w.r.t to \ch{CH3OH}.
\newline 
\noindent One should note that all observations discussed so far in this work measure gas-phase abundances of \ch{CH3OH}, \ch{CH2DOH}, and \ch{CH3OD}. Only \ch{CH3OH} has been detected in the ice of a few prestellar cores \cite{Boogert15}, mono-deuterated methanol has not. Therefore, the underlying assumption that abundances in the gas reflect the ice  \ch{CH2DOH}/\ch{CH3OD} ratio cannot be verified yet. 
\subsection{Possible routes to enhance deuterated methanol at later evolutionary stages}
\noindent D/H ratios derived from observations of the mono-deuterated isotopologues of methanol in hot corinos can be as high as a few percent \cite{Parise06survey,Taquet19}. This is higher than the values found in this work. Potential caveats of the network that could explain this discrepancy have been raised in Section 3.3. Additionally, modelled grain-surface abundances are compared to observed gas-phase abundances. It is unclear whether the D/H ratio of molecules observed in the gas phase in prestellar cores and in hot corinos are representative of the D/H ratio of the same molecule on the icy grains. Deuterium fractionation can also occur at temperatures as high as 70~K in the gas phase, which is caused by ions such as \ch{CH2D+} and \ch{C2HD+} and dependent species \cite{Millar89}. Therefore, the deuteration of some molecules can also take place after the prestellar stage. It cannot be ruled out that methanol is affected by this second, warmer deuteration channel.
\newline 
\newline 
\noindent Some high-mass star-forming regions, like Orion BN/KL, show a \ch{CH2DOH}/\ch{CH3OD} ratio that is close to unity \cite{Peng12}. Therefore, independent of processes that affect both mono-deuterated isotopologues, an additional formation mechanism for \ch{CH3OD} needs to take place at some stage of star formation.
\newline 
\noindent The hydroxy group in methanol can form hydrogen bonds with water ice \cite{Kawanowa04}. Spontaneous H-D exchange between (deuterated) methanol and (deuterated) water in the solid state has been proposed to change the abundances of hydroxy-deuterated methanol \cite{Ratajczak09}. This process does not occur for the methyl group. Starting from a mixture of \ch{CD3OD} and \ch{H2O}, rapid H-D exchange between the hydroxy group of methanol and water through H bonds has been observed in experiments upon ice heating to temperatures above 120~K through \ch{CD3OD} + \ch{H2O} $\longrightarrow$ \ch{CD3OH} + \ch{HDO}. This process has been modelled with a simplified rate equation model \cite{Faure15}. Results suggest that H-D exchange between hydroxy-deuterated methanol and either \ch{H2O} or HDO leads to the formation of methanol and subsequently stronger deuteration in water. Moreover, balancing through H-D exchange between the OH functional groups leads to an inverse scaling  of HDO/\ch{H2O} compared to \ch{CH2DOH}/\ch{CH3OD}.
\newline 
\noindent However, the underlying assumption of abundantly available hydroxy-deuterated methanol is taken as a given in the modelling and experimental work \cite{Ratajczak09,Faure15}, but not supported by experiments and observations of low-temperature environments, as well as models that simulate these parameter ranges. Moreover, these ideas do not account for the initial formation of \ch{CH3OD} on grain surfaces. 
\newline 
\noindent What has not been considered in modelling work yet is the interaction of \ch{CH3OH} with HDO and \ch{D2O}. H-D exchange between these two species has been studied with \ch{CH3OH} adsorbed on \ch{D2O} ice \cite{Kawanowa04}. Rapid and almost complete H-D exchange was observed in this experiment when the ice was warmed up to temperatures $>$~140~K. At these temperatures, the mobility of \ch{D2O} is enhanced. The H-D exchange also occurs if methanol is buried in \ch{D2O} ice, because it begins to travel to the ice surface at temperatures $>$~140~K before it desorbs at $\sim$~180~K \cite{Kawanowa04}. Rapid H-D exchange above 150~K and, therefore, \ch{CH3OD} formation at the cost of \ch{CH3OH} in \ch{D2O} ice has also been confirmed in other experiments \cite{Souda03}. Not all \ch{CH3OH} gets destroyed to form \ch{CH3OD}, some of it moves through the \ch{D2O} layers without undergoing exchange reactions and resides on top of the water ice until it desorbs. Exchange of H and D in the hydroxy group of methanol has also been seen in mixed ices consisting of equal monolayers of \ch{CH3OH} and \ch{CD3OD}, where efficient mixing occurs already at $\sim$~110~K \cite{Souda04}. This is very likely not representative of a realistic interstellar ice, but highlights the efficiency of H-D exchange. One can conclude that H-D exchange in the hydroxy group of methanol should be considered in astrochemical models of protostellar envelopes, where ices are expected to be heated. 
\newline 
\noindent Selective \ch{CH3OD} formation from \ch{CH3OH} would; however, depend on the availability of deuterated water in this case.
Water ice formation occurs predominately in the early stages of core formation \cite{vanDishoeck14}. Hence, methanol formation on top of water ice is a plausible scenario and has been reproduced with multi-phase models \cite{Taquet14}. Non-deuterated water forms primarily in molecular clouds, while its deuteration occurs later, during the dense core stage \cite{Furuya16}. Thus, HDO and \ch{D2O} are expected to be constituents of interstellar ices and could be available for H-D exchange with \ch{CH3OH}. HDO and \ch{D2O} are regularly detected in star-forming regions \cite{Vastel10,Coutens14,Jensen21obs}. Thus, HDO and \ch{D2O} should also likely be available in prestellar cores and part of the ice when methanol starts to form.
\newline 
\noindent H-D exchange could be a feasible way to enhance \ch{CH3OD} during the warm-up stage and explain why \ch{CH2DOH}/\ch{CH3OD} ratios are lower around protostars than the values suggested for prestellar cores. This requires extending the chemical networks presented here and calls for a physical model that accounts for the gravitational collapse of a prestellar core and the subsequent formation of a protostellar system. Such studies will be the topic of an upcoming project. Moreover, such a model can also test to what extent the second, warm deuteration channel influences the D/H ratios in methanol.
\FloatBarrier
\subsection{Comparison to a 3-Phase Model}
The deuteration of formaldehyde and methanol based on the abstraction scheme derived from the experiments by Hidaka et al. (2009) \cite{Nagaoka07,Hidaka09} has been explored in the past with a 3-phase chemical model (T12), which allows to trace the build-up of the mantle ice layer by layer \cite{Taquet12}. As in this work, a simple addition scheme to form deuterated species is tested. In order to properly assess the D/H ratio in formaldehyde and methanol, T12 starts the computations by solving a gas-phase network that sets the atomic D/H ratio and forms simple species, like CO. These steady-state abundances obtained by the core model are then used to compute the grain-surface chemistry. Therefore, in contrast to this work, T12 starts formaldehyde and methanol formation on the grain surface with CO already present; it does not need to be formed first. The physical parameter range explored in T12 is smaller than that considered in this work, a grid of low dust temperatures (8--12~K) and densities in the range of 10$^4$--5~$\times$~10$^6$~cm$^{-3}$ is investigated. The gas density of the fiducial model here (4~$\times$~10$^4$~cm$^{-3}$) lies at the lower range of the parameter space explored in T12, and the dust temperature range here is extended. Moreover, the diffusion energies in their model are higher, specifically, (0.5 - 0.8)~$\times$~E$_{\rm des}$ in comparison to the value of 0.3$\times$~E$_{\rm des}$ adopted here. Most importantly, T12 only considers activation energies, E$_{\rm A}$, of 400--1~400~K for reactions involving CO and formaldehyde. This is in contrast with this work, because in accordance with past modelling work \cite{Ruffle01,Garrod06,Drozdovskaya14}, a value of 2~500~K is adopted for the reaction with CO, and a range of 2~200--5~400~K for the reactions with \ch{H2CO}.  Nevertheless, the obtained results for the \ch{CH2DOH}/\ch{CH3OD} ratio obtained with the networks considering solely addition reactions are in agreement between T12 and Model 1 at 10~K. The results discussed with the fiducial model in Section 3.2 and Fig. 3 of T12 find the ratio to be close to unity across core ages of 10$^4$--10$^6$~yr. This still holds true when the parameter space explored by this work is expanded to cover the parameter range for E$_{\rm A}$ and n$_{\rm H}$ of T12 (Section 3.2.1, Table S1, and Fig. S1). The slight increase in the \ch{CH2DOH}/\ch{CH3OD} ratio at higher gas densities for older cores in this work in comparison to T12 predominantly stems from the inclusion of the formation of species with a higher degree of complexity than methanol in the chemical network of this work that take up some of the (deuterated) \ch{CH3O} for their formation. Overall, both approaches (2-phase and 3-phase models) lead to similar results.
\newline
\noindent The 3-phase model in T12 is also used to study the abstraction and substitution reactions towards the formation of (deuterated) formaldehyde and methanol. Layers that formed before conditions favored methanol formation and deuterium fractionation are buried deep in the ice mantle and likely not affected by the abstraction reactions. In contrast to the 2-phase approach in this work, 3-phase models usually account for processes like diffusion and photodissociation in the bulk ice and on the surface, and allow reactions in both ice components separately \cite{Kulterer20}. While the 2-phase model does not make the distinction between a chemically active bulk and surface, only the top two monolayers of the ice are chemically active. This ensures that not all of methanol (or any species) is considered to be taking part in grain-surface reactions. This explains the similarity of the results obtained by the 2-phase and the 3-phase model. Thus, both types of models can be used to trace the \ch{CH2DOH}/\ch{CH3OD} ratio as long as the number of chemically active layers in the 2-phase model is limited.
\newline 
\noindent Models 2A and 2B are predominately finding values for \ch{CH2DOH}/\ch{CH3OD} of 6--18 at 10~K, which agrees with the results of T12, but for higher densities and lower E$_{\rm A}$. The adaptations of Model 2A that include more up to date activation barriers for the reactions of (deuterated) formaldehyde with H or D (Section 2.2.4) were also used to cover the parameter space explored in T12 at 10~K. Model 2A$\_$\ch{CX2OX}, which adopts activation barriers from the \textsc{kida} database \cite{Wakelam12} for (deuterated) \ch{CH2OH} finds \ch{CH2DOH}/\ch{CH3OD} ratios $\sim$~10 for gas densities $\geq$~10$^6$~cm$^{-3}$, and for E$_{\rm A}$ in the range of 400 to 1~400~K for the CO + H $\longrightarrow$ HCO reaction (Fig. \ref{fgr:GCH2DOH_GCH3OD_CX2OX}). This is in good agreement with T12, which finds that densities of $\sim$~5~$\times$~10$^6$~cm$^{-3}$ are needed to obtain these \ch{CH2DOH}/\ch{CH3OD} ratios. Discrepancies between the 2-phase model and T12 arise at low gas densities. The ratios found here are higher than in T12, because the abundances of \ch{CH3OD} are much lower than for higher gas densities. This can be attributed to the fact that most \ch{CH3O} forms \ch{CH3OH}. In Model 2A$\_$\ch{CX2OX}, the formation of \ch{CH2OD} is suppressed due to its slower formation from H or D addition to \ch{H2CO} or HDCO, which is not taken into consideration in T12. If the E$_{\rm A}$ of the CO + H $\longrightarrow$ HCO reaction is set to 2~500~K, the \ch{CH2DOH}/\ch{CH3OD} ratio drops below unity for gas densities $\geq$~10$^6$~cm$^{-3}$ and to $\sim$~10 for gas densities below 10$^6$~cm$^{-3}$ in the 2-phase model. Note that this activation barrier has not been explored by T12. 
\newline 
\noindent If the barriers from quantum chemical calculations are adopted for all three production channels of \ch{X2CO} + X (Model 2A$\_$Song, Fig. \ref{fgr:GCH2DOH_GCH3OD_Song}), gas densities of $\geq$~10$^6$~cm$^{-3}$ combined with models that use E$_{\rm A}$~=~400--1~400K are still in agreement with T12. The differences between Model 2A$\_$Song and T12 that arise at lower densities follow the same explanation as for Model 2A$\_$\ch{CX2OX}.
\newline
\noindent In conclusion, both 2-phase and 3-phase models predict similar results for the \ch{CH2DOH}/\ch{CH3OD} ratio at prestellar stages. The preceding comparison yet again emphasizes that not the choice of the chemical model, but a careful exploration of the physical parameter space and uncertainties in the activation barriers of key reactions is more critical to predict the \ch{CH2DOH}/\ch{CH3OD} ratio in prestellar cores.
\section{Conclusions}
This work investigates possible formation routes of mono-deuterated methanol in prestellar cores. This is done with a simple addition-only model (Model 1), and a model incorporating the abstraction scheme derived from laboratory experiments \cite{Nagaoka07, Hidaka09}. This abstraction scheme and updates to the activation barriers of select key reactions are tested with Models 2A, 2A$\_$\ch{CX2OX}, and 2A$\_$Song. Model 2B investigates if boosting the relative rates derived from the experiments impacts the \ch{CH2DOH}/\ch{CH3OD} ratio. Based on this ratio, the question of the inheritance of mono-deuterated methanol from the prestellar stage to the protostellar stage is discussed. The most important findings are:
\begin{enumerate}
    \item Considering only simple H and D additions (Model 1) is neither sufficient to explain the lower limit for the \ch{CH2DOH}/\ch{CH3OD} ratio in prestellar cores, nor the observed ratios around low-mass protostars. 
    \item Models 2A and 2B, which implement the abstraction scheme of Hidaka et al. (2009) \cite{Hidaka09}, efficiently produce methyl-deuterated methanol at about one order of magnitude more than their hydroxy-deuterated counterparts at minimum. 
    \item The \ch{CH2DOH}/\ch{CH3OD} ratio should be $\geq$~10 in prestellar cores, which is in accordance with the lower limit of this ratio obtained from observations \cite{Bizzocchi14}. H-D exchange reactions should leave \ch{CH2DOH}, and therefore its D/H ratio, unchanged.
    \item \ch{CH3OD} formation at prestellar stages is inefficient, in agreement with the lack of observational detections of this isotopologue. At later evolutionary stages, H-D exchange reactions between HDO (and \ch{D2O}) and \ch{CH3OH} during the warm-up of ices might be forming \ch{CH3OD}. Thus, \ch{CH2DOH} could be inherited from the prestellar stage, but not \ch{CH3OD}.
    \item The \ch{CH2DOH}/\ch{CH3OD} ratio in warm environments of protostars is not inherited from the prestellar stage, but altered during the evolution from a prestellar core to a protostellar system. It is potentially a tracer of ice temperature: lower ($\sim$~10) for warm ices, and higher ($\gtrsim$~10) for cold ices.
    \item Variations of the modelled \ch{CH2DOH}/\ch{CH3OD} ratio are not only due to the considered physical environment, but also due to uncertainties in chemical parameters of two key reactions: CO + H $\longrightarrow$ HCO and \ch{H2CO} + H $\longrightarrow$ \ch{CH3O}, \ch{CH2OH}, HCO + \ch{H2}.  
\end{enumerate}
\noindent Due to the demonstrated efficiency of the production of methyl-deuterated methanol at prestellar core conditions, the \ch{CH2DOH}/\ch{CH3OH} or \ch{CHD2OH}/\ch{CH2DOH} ratios are potential means of constraining the natal cloud environment. This idea needs to be tested with more comprehensive chemical models and verified by dedicated observations.

\begin{acknowledgement}

BMK and MND acknowledge the Swiss National Science Foundation (SNSF) Ambizione grant no. 180079. MND also acknowledges the Center for Space and Habitability (CSH) Fellowship, and the IAU Gruber Foundation Fellowship. TJM and SA are grateful to the STFC for support through grant ST/P000312/1 and ST/T000198/1. C.W. acknowledges financial support from the University of Leeds, the Science and Technology Facilities Council, and UK Research and Innovation (grant numbers ST/T000287/1, and MR/T040726/1). We also want to thank the anonymous reviewers whose valuable comments helped to improve the quality of the manuscript.

\end{acknowledgement}

\begin{suppinfo}
The Supporting Information is available free of charge on the ACS Publication website under DOI:tbd and includes the following information in a PDF-file:
\begin{itemize}
    \item Description of formation pathways towards methanol and its deuterated isotopologues as obtained by Models 1, 2A, and 2B.
    \item Grain-surface \ch{CH2DOH}/\ch{CH3OD} ratios for the full parameter space exploration detailed in Section 3.2.1.
\end{itemize}

\end{suppinfo}

\bibliography{references}

\end{document}